\documentclass[10pt,letterpaper]{article}
\usepackage[top=0.85in,left=2.75in,footskip=0.75in]{geometry}

\usepackage{amsmath,amssymb}

\usepackage{changepage}

\usepackage[utf8x]{inputenc}

\usepackage{textcomp,marvosym}

\usepackage{cite}

\usepackage{nameref,hyperref}

\usepackage[right]{lineno}

\usepackage{microtype}
\DisableLigatures[f]{encoding = *, family = * }

\usepackage[table]{xcolor}

\usepackage{array}

\usepackage{algorithm,algorithmic}

\usepackage{xcolor}

\usepackage{soul}
\soulregister\eqref7
\soulregister\cite7
\soulregister\ref7
\soulregister\item7
\soulregister\textbf7
\soulregister\subsection7

\renewcommand{\hl}[1]{#1}

\newcolumntype{+}{!{\vrule width 2pt}}

\newlength\savedwidth

\raggedright
\usepackage[aboveskip=1pt,labelfont=bf,labelsep=period,justification=raggedright,singlelinecheck=off]{caption}

\makeatletter
\renewcommand{\@biblabel}[1]{\quad#1.}
\makeatother

\usepackage{lastpage,fancyhdr,graphicx}
\usepackage{epstopdf}
\fancyheadoffset[L]{2.25in}
\fancyfootoffset[L]{2.25in}
\begin{document}
\vspace*{0.2in}

\begin{flushleft}
{\Large
\textbf\newline{Addressless: A New Internet Server Model to Prevent Network Scanning} 
}
\newline
\\
Shanshan Hao\textsuperscript{1,2},
Renjie Liu\textsuperscript{1,2},
Zhe Weng\textsuperscript{1,2},
Deliang Chang\textsuperscript{1,2},
Congxiao Bao\textsuperscript{1},
Xing Li\textsuperscript{1,2*},
\\
\bigskip
\textbf{1} Institute for Network Sciences and Cyberspace, Tsinghua University, Beijing, China
\\
\textbf{2} Department of Electronic Engineering, Tsinghua University, Beijing, China
\\
\bigskip

%
%





* xing@cernet.edu.cn

\end{flushleft}
\section*{Abstract}
\hl{Eliminating unnecessary exposure is a principle of server security.}
The \hl{huge IPv6 address space enhances security by making scanning infeasible}, however, with recent advances of IPv6 scanning technologies, network scanning \hl{is again threatening server security}. 
In this paper, we propose a new model named addressless server, which separates the server into an entrance module and a main service module, and \hl{assigns} an IPv6 prefix instead of an IPv6 address to the main service module. 
The entrance module generates a legitimate IPv6 address under this prefix \hl{by encrypting the client address}, so that the client can access the main server \hl{on a destination address that is different in each connection}.
In this way, the model provides isolation to the main server, prevents network scanning, \hl{and minimizes exposure}. 
\hl{Moreover it provides a novel framework that supports flexible load balancing, high-availability, and other desirable features.
The model is simple and does not require any modification to the client or the network.}
We implement a prototype and experiments show that our model can prevent the main server from being scanned \hl{at a slight performance} cost.





\section*{Introduction}

\hl{Exhaustion of IPv4 addresses has long been recognized and is now a reality.}
\hl{IPv6\cite{rfc8200} was proposed in 1995 to solve this problem.}
The main \hl{improvement} is the 128-bit address over the 32-bit IPv4 address, \hl{together with other goals like end-to-end feature and better security.}
As of March 2020, about 25\% information resources ( websites, emails, etc.), 60\% DNS servers, and 30\% Internet clients support IPv6 \footnote{https://www.vyncke.org/ipv6status/}. 
With the rise of the Internet of Things, 5G, and cloud computing, it is predicted that more than 75 billion devices will be connected to the Internet by 2025 \footnote{https://www.statista.com/statistics/471264/iot-number-of-connected-devices-worldwide}, while there are only 4 billion IPv4 addresses in total.
\hl{An inevitable and faster adoption of IPv6 can be expected.
IT giants such as Facebook\footnote{https://www.facebook.com/ipv6/} and Microsoft\footnote{https://teamarin.net/2019/04/03/microsoft-works-toward-ipv6-only-single-stack-network/} have been moving to an IPv6-only internal network.
An} IAB (Internet Architecture Board) statement expected that the IETF would stop requiring IPv4 compatibility in new or extended protocols, and future work would optimize for and depend on IPv6~\footnote{https://www.iab.org/2016/11/07/iab-statement-on-ipv6/}. \hl{Pure} IPv6 is the future of the Internet.




\hl{IPv6 is developed with security in mind, realizing that security as well as privacy has always been one of the biggest threats to the Internet.} 
\hl{It is natural to start with utilizing the massive address space of IPv6 to enhance security and privacy, considering that this is its biggest difference from IPv4.}

\hl{By making network scanning difficult, this massive address space of IPv6 has naturally provided preliminary protection. 
The IP address is the identifier and locator of the Internet. 
Thus network scanning is usually the first phase of an attack to obtain the IP addresses of potential victims. 
This is easy in IPv4.} The scanning time of the entire IPv4 address space is only about 45 minutes\cite{zmap}. However, it takes more than 100 quintillion years to scan the entire IPv6 address space with the same efficiency.
\hl{The massive address space of IPv6 helps a lot to hide the address of a device.}


\hl{However, on one hand, NAT in the IPv4 era, albeit not designed for security, brings the byproduct of hiding devices and topology of the network from the outside.
NAT is stateful and violates the end-to-end principle, thus deprecated by IPv6 designers. 
Many network administrators, however, are so used to the invisibility provided by NAT that they feel that NAT-less IPv6 rather poses a threat of exposure, despite the difficulty of scanning IPv6's huge address space.}

\hl{On the other hand, recent advances of IPv6 scanning technology have threatened the preliminary invisibility provided by the huge address space.
Various approaches have been proposed} to scan the IPv6 Internet \hl{more efficiently mainly in two ways:} collecting active IPv6 address records\cite{b21, b22, b23, b24, b25, b25-2, b26, b26-2}, and using statistical and machine learning methods to generate hitlists\cite{b27, b28, b29, b30, b31, b32-dp}. 
Most of these approaches are server-specific. 
Scanning IPv6 clients remains difficult, which brings many security benefits for the clients. 

\hl{We cannot help but think, \textit{can we further utilize the IPv6 address space to make IPv6 servers unscannable as well, avoiding as much unnecessary exposure as possible, thereby allowing servers to enjoy the security advantages brought by invisibility, but not violating end-to-end reachability?}}


\hl{There have been attempts to enhance the anti-scanning feature of the IP addresses, mainly by: 1) generating addresses that is semantically opaque and more random\cite{rfc7217}, 2) using temporal\cite{rfc4941, rfc7707} or hopping\cite{hopping-2005, hopping-slaac, Dunlop-2012, b34-IoT-2017, MTD-2012, FRVM-2018} addresses with short lifetime, and further 3) using a unique address for each connection\cite{per-connection-2001, per-connection-idea, protocol-stack-virtualization, ota-07, mix_ISP} or even packet\cite{per-packet-08, per-packet-16}.

The former two approaches are basically incremental improvements.
Each device still has one IP at a time which can be scanned.
The per-connection address is an interesting idea but previous models are very complex, unscalable, and hard to deploy.
Specifically, they largely remain within the framework of dynamically shared address pools. Address mapping and address collision become problematic, the network needs to be modified to enable routing, and the system is complicated. 
Moreover, few works are about servers, and generally require synchronous cooperation of the client-side.}

\hl{On the other hand,} the current network security model relies on encryption. \hl{For instance, SEND\cite{send1}, TLS\cite{tls}, and DNSSEC\cite{dnssec} are used at the data link layer, the transport layer, and the application layer, respectively.}
At the network layer, IPsec\cite{ipsec} \hl{encrypts the payload of IP packets but the (outer) IP addresses are left exposed.} 
Since TCP/IP was introduced, the creators of the Internet have considered introducing encryption into the IP address itself. \hl{This is an extravagant dream in the era of IPv4 since addresses are scarce and reused with great care}. However, this is changed in IPv6, and the 128-bit address provides sufficient space to carry the encrypted information. 

\hl{Previous work\cite{per-packet-08, mix_ISP, per-packet-16} propose to encrypt host identity into pseudo-random temporal addresses, so that the host identity is hidden from the outside network. But it has to be done on local routers, otherwise routing will fail. The network needs to be modified and it is sort of encapsulation. 
CGA\cite{rfc3972} applies hashing in auto-generation of addresses, but it aims to solve link-local address spoofing and again the addresses are exposed.
The literature has not seen a simple and scalable mechanism to introduce encryption into the IP address itself. } 


In this paper, a model named addressless public server is proposed.
\hl{We creatively use the prefix delegation mechanism\cite{dhcpv6} in a way that is different from its original intention. 
So that we introduce encryption into the per-connection address in a very simple way and no modification of other participants is needed.
And we make it usable for public servers to make them difficult to be scanned. 
Not only is security strengthened in this way, but our model also offers a novel architecture to enable flexible high-availability and other features.}

The addressless sever has two modules, one module provides independent entrance of the server, the other provides the main services. 
When receiving access requests, the entrance module generates a destination address using encryption, and redirects the request to the generated destination address. 
\hl{This destination address is different for each connection request.}
The main service module is allocated a prefix instead of an IPv6 address and listens on all the addresses under the prefix.
When the main service module receives a data flow, it conducts verification on the \hl{destination address}, and only responds to flows that pass the verification.
\hl{Scanning traffic and attackers that visit the main service module directly cannot pass the verification, therefore} will be immediately dropped. 
In this way, we make the main server imperceptible.
 
This model separates the network entrance and provides isolation for the main server. 
It allocates a prefix instead of an address to the main server, so that it no longer has \hl{one address at a time, thus} imperceptible by the outside network. 
At the same time, our model \hl{naturally supports} flexible high-availability solutions such as lightweight load balancing, \hl{active-active cluster, and CDN. More security and functional mechanisms can be developed based on it. All in all,} our model provides a novel perspective on various problems faced by public servers.

\section*{Background and Related Work}
Our model uses the massive IPv6 address space to prevent the server from being scanned. So in this section, \hl{we first introduce network scanning and IPv6 address space security. Then we introduce prefix delegation, the address configuration mechanism that our model built upon. Finally, we make a detailed analysis of previous work in using IPv6 address space to enhance security, and compare our model to the most related ones.}
 
\subsection*{Network Scanning}
Network scanning is a technology to collect the active addresses by sending packets to a \hl{huge} set of addresses. 
\hl{Early scanning techniques like} Nmap\cite{nmap} often take days to scan the entire IPv4 address space. 
Zmap\cite{zmap} proposed in 2013 reduces the scanning time of the entire IPv4 address space to about 45 minutes. This makes the cost of scanning under IPv4 negligible. 

Due to the massive address space, network scanning in IPv6 has always been considered impossible. 
Although Zmapv6\cite{zmapv6} is proposed to scan IPv6 addresses, it is only a scanning tool \hl{that uses the same technique as Zmap without improving} scanning efficiency. 
However, a series of \hl{techniques} has been \hl{proposed} in recent years to reduce the difficulty of IPv6 scanning. 
 
The work on IPv6 scanning can be divided into two types. One is to generate hitlists by collecting active addresses on the Internet. Various sources can be used to collect active addresses. 
Fiebig et al.\cite{b21} propose to generate the hitlist using DNS data, and Borgolte et al.\cite{b22} propose to generate the hitlist using DNSSEC-signed reverse zones, while rDNS data is used by Fiebig et al.\cite{b23}. 
These researches mainly focus on generating the hitlists of Internet servers, while Beverly et al.\cite{b24} and Rohrer et al.\cite{b25} introduce approaches to collect router addresses\hl{, and Rye et al.\cite{b25-2} probe and collect addresses of last-hop routers using traceroutes.}
Gasser et al.\cite{b26} summarize these methods and give a large hitlist set\hl{, and publish a compiled, open-source, and frequently updated hitlist whose quality is enhanced using active measurements\cite{b26-2}.}

The other is to generate hitlists by predicting active addresses using statistical or machine learning algorithms. 
Foremski et al.\cite{b27} first propose that active IPv6 addresses can be predicted by statistical algorithms. They introduce Entropy/IP, an algorithm to generate hitlists using the Bayesian algorithm. 
Ullrich et al.\cite{b28} introduce a scanning algorithm based on pattern recognition, and Zuo et al.\cite{b29} analyze the active addresses and make predictions through association rule learning. 
Murdock et al.\cite{b30} propose 6Gen, an algorithm that uses some active addresses as seeds to generate hitlists. 
Liu et al.\cite{b31} introduce 6Tree, which uses hierarchical clustering to predict addresses and generate hitlists. 
\hl{Deep learning methods have also been introduced. Cui et al.\cite{b32-dp} stack gated convolutional network to encode address structure and generate hitlists.}

\subsection*{IPv6 Address Space Security}
Since the birth of IPv6, researchers have been looking for ways to increase the security and privacy of IPv6 addresses.

In RFC 4291\cite{rfc4291}, an IPv6 address is divided into two parts: a 64-bit prefix and a 64-bit identifier interface (IID). 
The most widely used IPv6 address \hl{configuration} methods are SLAAC\cite{slaac} and DHCPv6\cite{dhcpv6}. 
In SLAAC, the IID is generated \hl{from} the MAC address of the device using the EUI-64 algorithm\cite{rfc2464}. 
This makes the IID part of the IPv6 address remaining \hl{constant in the lifetime of a device}. 
\hl{In DHCPv6, the rules of address configuration are determined by the network administrator.} 
In the early days, \hl{these rules are simple and regular, which make the addresses used in the network showing} obvious patterns. 
Scanning is quite easy for both SLAAC and DHCPv6 addresses, thus posing a threat to security.

To solve this problem, SLAAC Privacy Extension (SLAAC PE) is proposed in RFC 4941\cite{rfc4941}. 
In RFC 4941, the client is recommended to use a temporary address to communicate with servers. 
The IID of the temporary address is one-time, generated \hl{by applying} the MD5 algorithm\cite{md5} to the previous IID. 
In this way, the security and privacy of the client are enhanced in SLAAC.
DHCPv6 also proposes an approach of allocating temporary addresses\cite{b12}, but it is not widely used because it brings a series of problems\cite{b13}.

More in-depth work is introduced on this basis. 
Semantically opaque IID is recommended to be used in SLAAC and DHCPv6 by RFC 7217\cite{rfc7217}, RFC 7493\cite{rfc7943}, and RFC 8065\cite{rfc8065}, etc.
RFC 7707\cite{rfc7707} recommends to reduce the lifetime and increase the randomness of IPv6 addresses used by network devices based on a summary of scanning algorithms.

At the same time, researchers also put forward some work on the \hl{measurement of active} IPv6 address space. 
Plonka et al.\cite{b19} analyze the temporal and spatial characteristics of active IPv6 addresses.
Li et al.\cite{b20} describe the distribution characteristics of Internet IPv6 prefixes. 
These works demonstrate how the IPv6 addresses are actually used thus objectively shows the security status of the IPv6 address space.

There are also a few works focused on the detection and defense of IPv6 scanning.
Fukuda et al.\cite{b32} introduce an approach to detect IPv6 scanning and evaluate relevant severity. 
Plonka et al.\cite{b33} introduce kIP, a new approach to increase the anonymity of IPv6 addresses.

\subsection*{IPv6 Prefix Delegation}
In our model, the main service module is assigned an IPv6 prefix. 
\hl{Assigning a prefix to a device is allowed in IPv6,} typically using the DHCP-PD\cite{dhcpv6} mechanism. 
DHCP-PD is originally proposed to allow a DHCP server to assign a prefix to a DHCP client, so that this DHCP client can further allocate the addresses under the prefix to other devices. 
\hl{In our model DHCP-PD is used for a different purpose.} 
The main service module is assigned a prefix using DHCP-PD, but after that, it uses the addresses under the prefix itself, instead of allocating them to others. 

 
Besides DHCP-PD, prefix delegation is also used in other scenarios. 
RFC 8273\cite{rfc8273} \hl{allows each device in the same subnet} to be configured with a unique IPv6 prefix, so that \hl{they are logically under different subnets and} cannot send packets to each other except through the first-hop router.
Using this, isolation is provided between the devices in a shared-access network. 
However, \hl{each device only uses one fixed address under the prefix.} \hl{To the best of our knowledge, the literature has not made deeper use of the prefix allocated to a device.}


\subsection*{\hl{Using IPv6 Address Space to Enhance Security and Privacy}}

\hl{\textbf{IP address hopping.} The technique to dynamically and frequently change the IP address of a device has long been used to prevent attackers from finding the target since the era of IPv4.
In the IPv4 era, it can only be achieved with the help of the network service, such as DHCP servers\cite{hopping-2007} and SDN\cite{MTD-2012,FRVM-2018}. 
In the IPv6 era, the massive address space and SLAAC enable auto-configuration of dynamic addresses\cite{hopping-slaac}. 
IP address hopping of a server is harder, since its address needs to be known by clients to allow inbound traffic. 
In previous work, the sender has to update the addresses synchronously with the receiver based on a shared secret\cite{hopping-2005, Dunlop-2012, b34-IoT-2017}.

\textbf{Per-connection address and beyond.} 
IPv6 addressing model specifically supports assigning multiple IP addresses to a single interface\cite{rfc4291}. 
Thus researchers have gone beyond address hopping to assign each connection (or a set of closely related connections) a unique address\cite{per-connection-2001, per-connection-idea, protocol-stack-virtualization, ota-07, mix_ISP} to enhance privacy. Our model also uses per-connection addresses, and a detailed comparison is given later in this subsection.

The per-connection address is not the end of the road either. 
Researchers extend this idea temporally to propose per-packet address\cite{per-packet-08, per-packet-16}, and spatially to propose prefix alteration\cite{prefix-alteration-2012, prefix-alteration-2019, prefix-alteration-mobile}. The per-packet address means to use a unique address for each packet. However, the stronger privacy is achieved at the cost of higher complexity of demultiplexing packets to flows and modification of local networks to enable routing\cite{per-packet-08, per-packet-16}. Prefix alteration means not only the IID part but also the prefix part of a device's address can be varying. It can be achieved by prefix hopping, prefix bouquets, prefix sharing, and variable-length prefix\cite{prefix-alteration-2012, prefix-alteration-2019}, or by exploiting mobile IPv6\cite{prefix-alteration-mobile}.

\textbf{Limitation of previous work.}
The idea of using address space to hide node identity stems from IP address hopping in the IPv4 era, which can only be achieved through a pool of dynamic addresses shared by a group of devices due to scarcity of addresses. 
Although previous works have applied the idea to IPv6 and extended it to per-connection or even per-packet address, in essence, they have not gone beyond dynamically shared addresses pool. 
That is, devices are still sharing a set of dynamic addresses, although this set becomes enormous; it contains all the addresses under the IPv6 prefix.

The scheme of dynamically shared addresses pool brings two problems. First, mapping an address to a device or a connection and the related routing are difficult. 
Early work is stateful, the mapping needs to be recorded\cite{per-connection-2001, protocol-stack-virtualization, ota-07}. 
More recent work achieve stateless mapping by encrypting the host identity into the one-time pseudo-random address. 
However, decryption needs to be done at the local router because the host identity needs to be extracted for local routing\cite{per-packet-08, mix_ISP, per-packet-16}. Modification of the network service makes it unlikely to be deployed.} 

\hl{Second, IPv6 enables self-generated pseudo-random addresses without relying on stateful DHCP servers. But making sharing of addresses stateless brings the possibility of address collisions.
Previous works skip this problem by stating that the possibility is negligible or use duplicate address detection to detect and avoid collisions\cite{protocol-stack-virtualization, ota-07}. Some use external unique identifiers to generate addresses, such as Electronic Product Code\cite{EPC} in IoT scenarios. 

Further, few works are about servers. On one hand, it is complex for a server to maintain a huge set of addresses assigned to each of its connection under this dynamically shared addresses pool scheme. For instance, Sakurai et al.\cite{ota-07} use sliding windows to maintain active addresses. On the other hand, the contradiction between inbound traffic and dynamic addresses has not been solved. Sender and receiver need to cooperate and synchronously update a pseudo-random sequence of addresses calculated based on a secret\cite{hopping-2005, Dunlop-2012, b34-IoT-2017, ota-07}. This is infeasible for public servers.

\textbf{Comparison of our model.}
We innovatively utilize the prefix delegation mechanism\cite{dhcpv6}.
Our model is built on the idea of the per-connection address. 
However, the device is assigned all the addresses under the prefix, thus the complexity of routing and address collision is completely eradicated, and modification of the network service is no longer necessary. 
And specifically for a server with a huge amount of connections, the maintenance of the active addresses is no longer a problem.

Prefix delegation also enables us to introduce encryption into the address at the endpoint. 
Previously it has to be done on local routers\cite{per-packet-08, mix_ISP, per-packet-16} otherwise the dynamically shared addresses cannot be routed properly. 
Together with a novel stateless salting algorithm, we achieve stateless mapping of the per-connection address to the client.

And to make the public server accessible at the per-connection addresses, we innovatively separate the Internet entrance module from the main service module, and use the entrance module to generate the encrypted address and redirect the client. 
In this way, no modification of the client-side is needed. 
The exposed entrance module bears no other logic and is simple, and the main service is isolated and hidden in the huge address space.}

\section*{Design of Addressless Server}
In this section, the design of addressless server is presented.

\subsection*{Design Principles}
The addressless server model is designed to prevent the server from being perceived and scanned by the attackers. To reach this goal, the model takes advantages of the following features:
\begin{enumerate}
\item Separate the Internet entrance module from the main service module, and provide isolation to the main service module.
\item Allocate a prefix instead of an address to the main service module.
\item Eliminate the one-to-one correspondence between the server and the IP address.
\item Introduce encryption into IPv6 Address to make use of the redundant IPv6 address space. 
\end{enumerate}
In this way, the model can provide a triple guarantee for the server: strip the Internet entrance from the main server, hide the legitimate address in the massive IPv6 address space, and use encryption to ensure that only the legitimate address generated by the entrance module can be visited. By decoupling the server and the IP address, attackers can no longer use the IP address as the identification of the server. This is the meaning of ``addressless’’. Through this, the server is protected from being perceived and scanned by the outside devices, so that security is enhanced.

\subsection*{System Design}
\begin{figure}[!h]
\centering
\includegraphics[width=12.5cm]{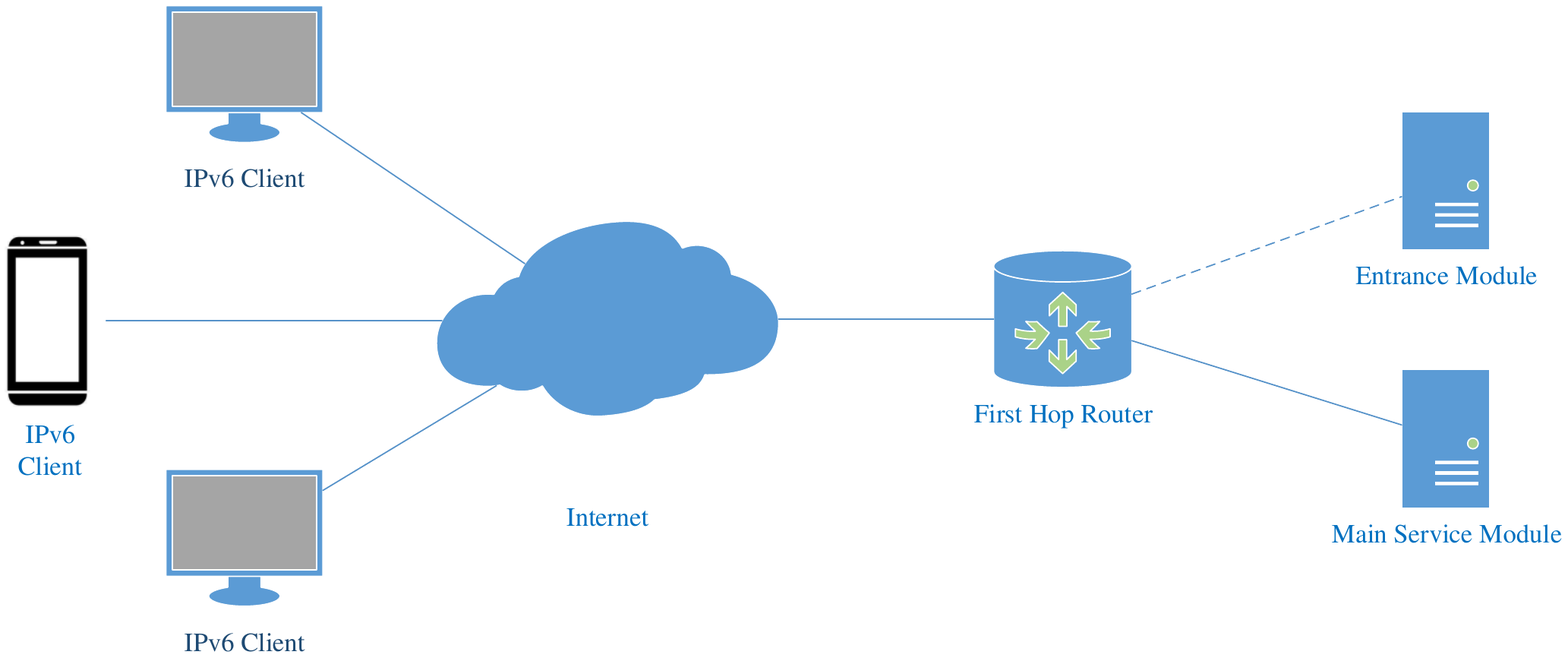}
\caption{{\bf Topology of the Addressless Server.}
\hl{The server is separated into the entrance module and the main service module. The main service module is directly connected to the first-hop router.} The dotted line \hl{indicates} that the entrance module can be deployed anywhere on the Internet.}
\label{Fig 1}
\end{figure}

We divide the server into two modules, the entrance module and the main service module. The entrance module has a fixed Internet address; the main service module is configured with a prefix and uses all the addresses under the prefix to communicate with clients. The topology of the addressless server is shown in Fig~\ref{Fig 1}. 

The main service module is directly connected to the first-hop router. The main service module and the first-hop router are both configured with a non-public IPv6 address. This address is used for prefix delegation and routing. It is usually a link-local address. If there are more management requirements, other private addresses such as ULA\cite{ula} can also be configured. The first-hop router routes all the packets whose destination addresses are under \hl{this} prefix to the main service module. The entrance module is configured with a fixed IPv6 address. This address should be configured as an AAAA record in the DNS system. 
 
When an Internet \hl{client} initiates a connection with the server, it first sends a DNS query to the DNS server. The DNS server returns the entrance address to the client. Then the client sends the request to this address. After receiving the request, the entrance module uses the prefix of the main service module and the source address of the packet to calculate an IPv6 address through an encryption algorithm, then returns this address to the client. Finally, the client initiates connections with the main service module using this address as the destination address. 
 
The calculation process can be formulated as the following equations. We denote the client address as $SA$, the encryption process as function $f()$ (the specific process of $f()$ is discussed in next subsection), and the prefix of the main service module as $prefix$, then the destination address is:

\begin{equation}DA_{1:N} = prefix\label{eq1}\end{equation}

\begin{equation}DA_{N+1:128} = f(SA)\label{eq2}\end{equation}

\hl{To achieve compatibility, clients that are agnostic of our model should be able to communicate with the public server without modification. So we use the redirection mechanism to achieve the process of entrance module returning the generated address and client connecting that address.} In this case, the entrance module temporarily redirects the client’s request to the generated address. After receiving this message, the client sends a new request to the main service module. When the main service module receives a flow, it verifies the destination address using the source address through the same encryption algorithm. The verification process can be described by Eq~(\ref{eq3})

\begin{equation}Res = g(SA, DA_{N+1:128}))\label{eq3}\end{equation}

In Eq~(\ref{eq3}), $g()$ is the verification function. $Res$ is a bool value. True means the flow passes the verification while False means the flow fails the verification. If the verification fails, the server discards the packets. 
 
\begin{figure}[!ht]
\centering
\includegraphics[width=12.5cm]{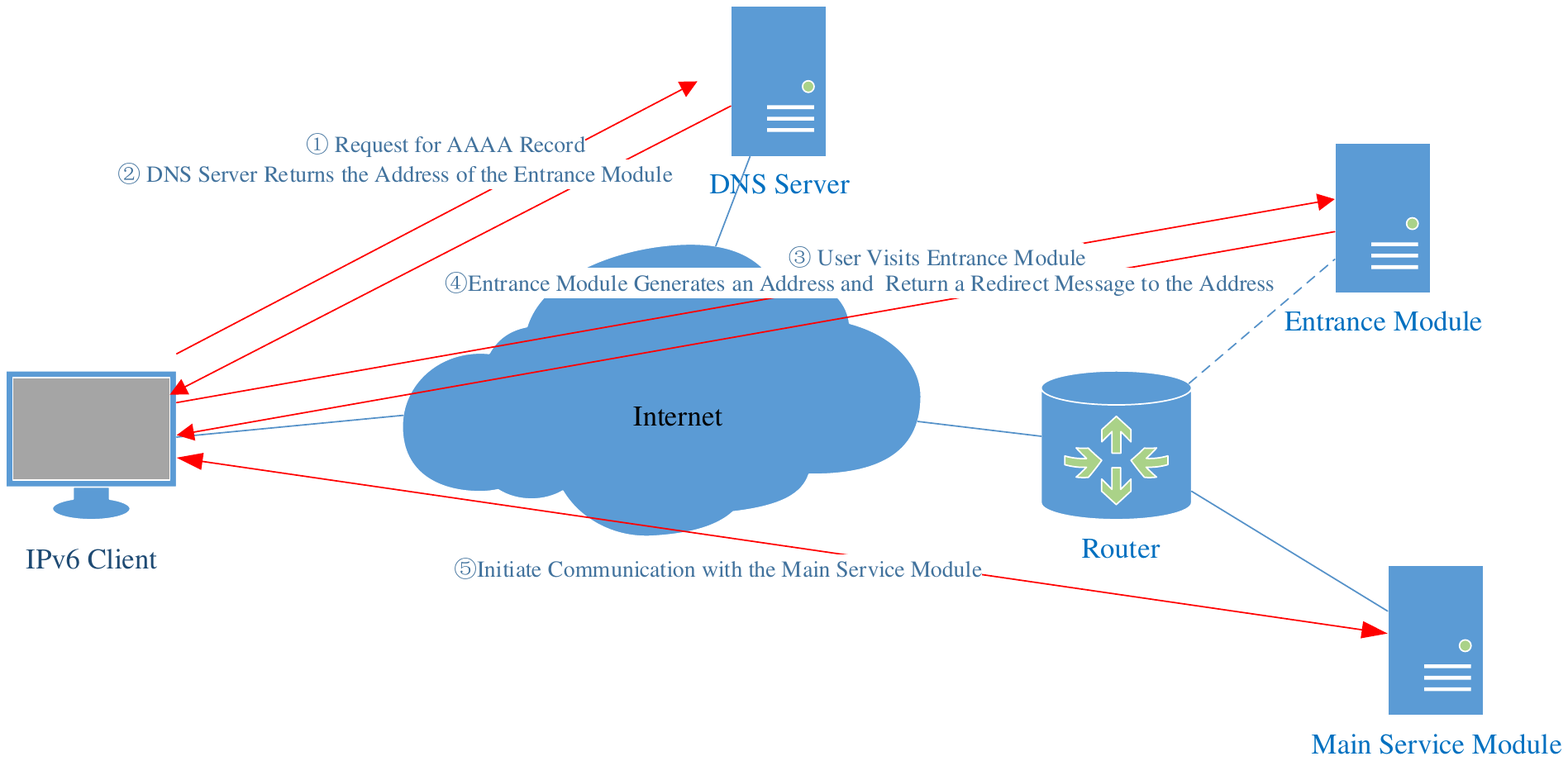}
\caption{{\bf Mechanism of the Addressless Server.} \hl{The communication process of how a client successfully initiates a flow to the server.}}
\label{Fig 2}
\end{figure}

The mechanism is described in Fig~\ref{Fig 2}. 

Noted that the client here is just a common IPv6 client. It is configured with an IPv6 address, not a prefix. The address used in the whole connection lifetime should remain unchanged. Otherwise, it may cause the source address used in the encryption and the source address used in the decryption different, leading to the failure of the verification. 
 
The verification process is performed in each flow. The main service module only needs to perform the verification once for each flow because the source address and destination address \hl{remain unchanged during the flow}. \hl{Once} the connection is established, the server can directly accept the following packets until the end of the flow. \hl{After} a connection is terminated, the client should visit the address of the entrance module if it is going to initiate another connection, and the entrance module generates another address and redirects the request to this new address. 
 
To prevent attackers from intercepting data flows and launching replay attacks, the generated address should be different each time. To reach this goal, we add a time-varying factor in the encryption process $f()$, which is described as salt. In this case, when the attacker intercepts the \hl{packets} and forges an attack message using the source and destination address \hl{from those packets}, it will not be effective because the legitimate destination address is different because the salt \hl{has changed}. 
 
In our model, the entrance module can be deployed anywhere on the Internet, \hl{logically and geographically}. It can be configured on the same device as the main service module, and connected to the Internet through the same first-hop router. The entrance module can also be configured on different devices in the same subnet, or even a location far from the main server on the Internet. The address allocated to the entrance module can be one of the addresses under the prefix delegated to the main service module, or it can be a totally independent global unicast address. Furthermore, considering that the entrance module only provides simple and reproducible services, it is easy to stack or distributed deploy the entrance modules. All in all, the configuration of the entrance module is very flexible. Various strategies can be used on the deployment of the entrance modules to achieve better results. 
 
In the addressless server mechanism, the entrance module is responsible for calculating the destination address and \hl{returning} a redirect message to the client. If the server needs a stricter strategy, \hl{the entrance module} can also provide user authentication service, and only replies to the client who passes the authentication. This ensures that all the clients that can perceive the main \hl{service module} are authenticated, which further ensures server security without affecting the simplicity. 
 
From the above discussion, we can see that our model takes advantage of the redundant space of the IPv6 address by introducing encryption into IPv6 suffixes. The IPv6 space is very large, and only a small part is actually used in traditional scenarios. As a result, we can `waste’ the address space to prevent the server from being scanned. We assign a prefix to the server to free up the suffix space and let it carry the authentication information. In our model, the use of the prefix and the suffix of the destination address is different. The prefix is used for routing while the suffix is used for carrying authentication information to provide additional security benefits.

\subsection*{Encryption Algorithm}
The encryption in this paper is essentially a signature-verification process. When a client initiates communication, the entrance module first signs the source address, embeds the result into the destination address, and returns it to the client. Then the client initiates connections to that address, and the main service module conducts verification using the source address and the destination address. Since we do not need to restore the message, $f()$ here does not need to be reversible. 
 
The encryption process $f()$ is described as follows:
 
\begin{equation}H\_SA=Hash(SA)\label{eq4}\end{equation}
 
\begin{equation}P\_SA=\Phi(H\_SA, salt)\label{eq5}\end{equation}
 
\begin{equation}DA_{65-128}=e(P\_SA,key)\label{eq6}\end{equation}
 
\begin{equation}DA=strcat(prefix,DA_{65-128})\label{eq7}\end{equation}
 
And the verification process $g()$ is described as follows:
 
\begin{equation}H\_SA=Hash(SA)\label{eq8}\end{equation}
 
\begin{equation}P\_SA=e^{-1}(DA_{65-128}, key)\label{eq9}\end{equation}
 
\begin{equation}Result=\Psi(P\_SA,H\_SA,salt)\label{eq10}\end{equation}
 
In Eq~(\ref{eq4})-Eq~(\ref{eq10}), $SA$ is the client address (source address of the connection \hl{request}); $DA$ is the generated address under the prefix of the main service module \hl{(destination address of the connection request)}. 
 
In Eq~(\ref{eq4}) and Eq~(\ref{eq8}), $Hash()$ is a hash function to convert the source address into a 64-bit sequence. A hash function guarantees that the sequence is uniformly distributed in the entire \hl{range} space. Any hash function is feasible here, \hl{including cryptographic hash functions such as md5\cite{md5} and SHA-128\cite{sha}, or string hash functions such as DJB and BDKR}. The string hash function is a better choice here because of the higher efficiency. DJB algorithm is used in our prototype.

In Eq~(\ref{eq5}) and Eq~(\ref{eq10}), $salt$ is used as the time-varying factor. We add the salt in function $\Phi()$, which is discussed in the next subsection. Function $\Psi()$ is the verification function determined by $\Phi()$.
 
In Eq~(\ref{eq6}) and Eq~(\ref{eq9}), $e()$ is the encryption function, and $e^{-1}()$ is the decryption function. $key$ is the encryption key.
 
\begin{figure}[!ht]
\centering
\includegraphics[width=12.5cm]{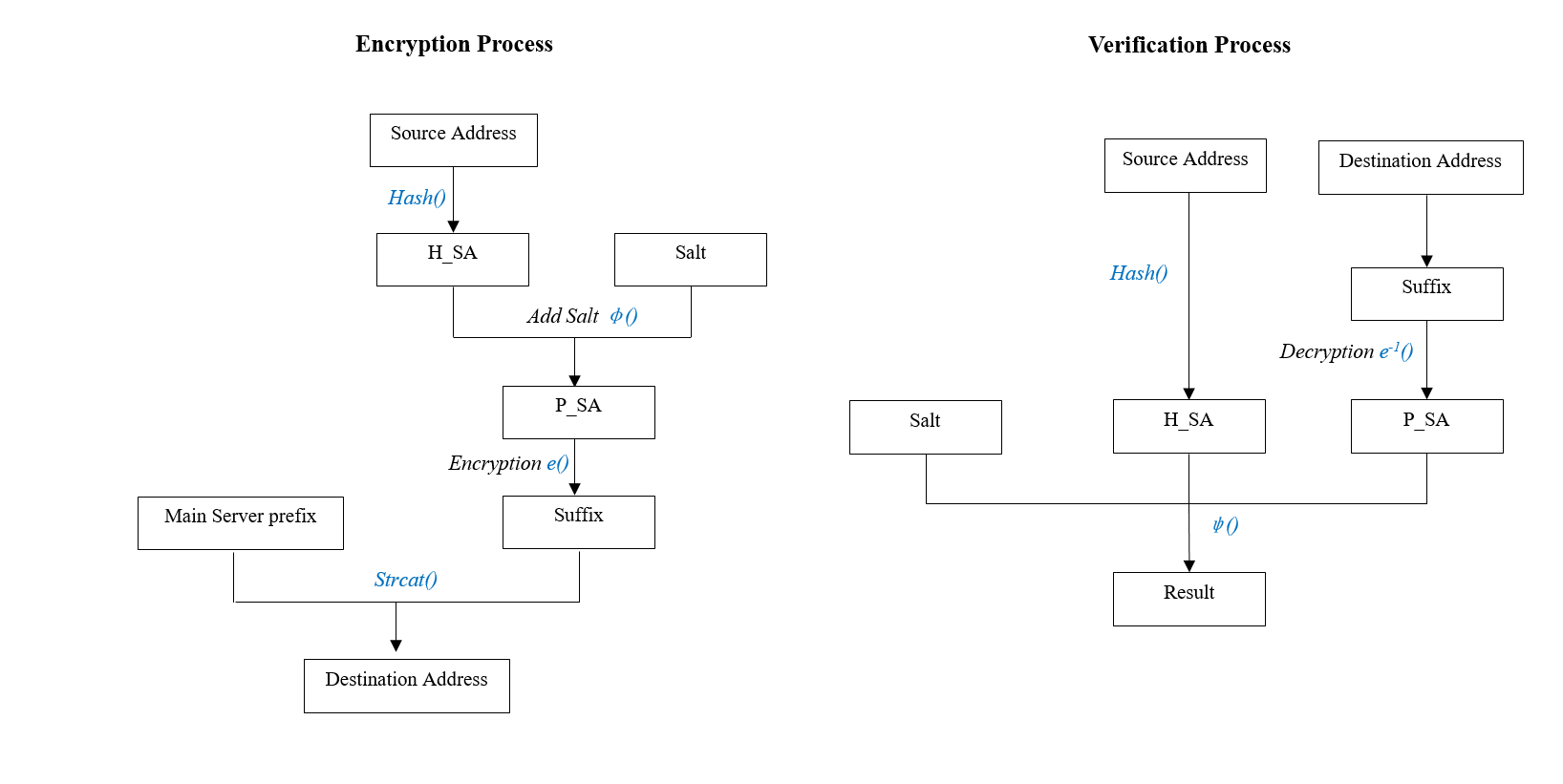}
\caption{{\bf Encryption Process and Verification Process.}}
\label{Fig 3}
\end{figure}

Fig~\ref{Fig 3} shows the encryption and verification process briefly.

Theoretically, the encryption algorithm $e()$ can be arbitrary\hl{, as long as the ciphertext can be held in the suffix space in some way}.
However, we should consider it from two perspectives: security and efficiency. 
Considering security, a mainstream encryption algorithm should be used here. 
\hl{There are two categories of encryption algorithms: the symmetric encryption algorithm and the asymmetric encryption algorithm. 
The symmetric encryption algorithm has the following advantages: to achieve the same security level, it has shorter ciphertext length and key length. 
While the advantage of the asymmetric encryption algorithm is that it allows the public key not to be secret. 
In our model, the key is used only by the entrance module and the main server module, which are both under the control of the service owner. 
There is no need to distribute the public key, and it is not a challenge to keep the encryption key secret, thus there is no need for the asymmetric encryption algorithm. 
Considering efficiency, the symmetric encryption algorithm is also better for faster encryption and decryption processes.
As a result, the symmetric encryption algorithm is a better choice in our model, which can ensure a faster encryption/decryption process and a higher level of security for a given ciphertext length.}
 
The most widely used symmetric encryption algorithms are DES\cite{des}, 3DES\cite{des}, AES\cite{aes}, etc. 
Considering the 64-bit length of the IPv6 suffix, to make the encryption easier, DES or 3DES is better here. 
Although DES is often considered insecure in the modern network environment, it can provide a sufficient level of security in our scenario (discussed in the Security Analysis Section) while it is much faster than 3DES. 
As a result, we use DES as $e()$ in our prototype implementation. 
\hl{Nevertheless, our model does not place restrictions on exactly which encryption algorithm is used; that is up to the choice of the server owner.}

\subsection*{Salting Algorithm}
We discuss the generation of the time-varying factor (salt) in this subsection. To make the generated addresses unpredictable and the replay attacks ineffective, adding salt is necessary in the address generation. 

\begin{figure}[!ht]
\centering
\includegraphics[width=12.5cm]{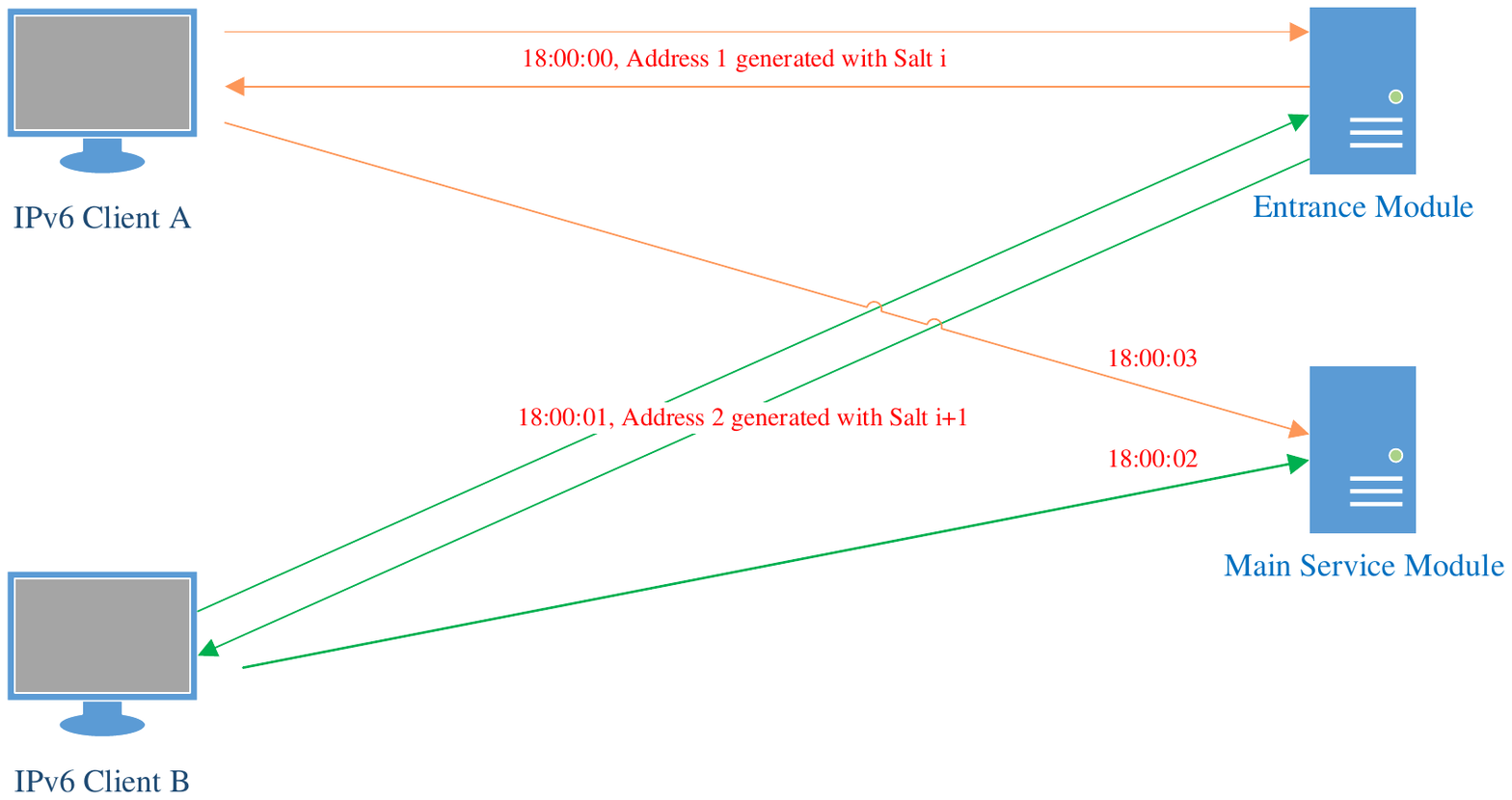}
\caption{{\bf Stateful Salt is \hl{NOT} a Good Choice.} \hl{Client B visits the entrance module later than client A and gets an address generated with salt i+1. Due to different delay, B visits the main service module earlier than A and fails verification because the main service module is expecting salt i. }}
\label{Fig 4}
\end{figure}

\hl{Stateful salt is the common choice in many salting scenarios. First, we consider if we can use stateful salt in our model.} In this case, the entrance module and the main service module save the same state sequence. When an address is generated by the entrance module, it uses the current state as the salt, then hops to the next state. It is similar in the verification process in the main service module. However, it is not a good choice here because synchronization is a challenge. As shown in Fig~\ref{Fig 4}, \hl{first} client A visits the entrance module and obtains the redirect address $Address_1$ generated with salt i, \hl{later} client B obtains the address $Address_2$ generated with salt i+1\hl{. But because of different delay}, B visits the main service module earlier\hl{. At this time, the state in the main service module is still the salt i, thus the packets of B will fail verification and get wrongly dropped. This situation will happen frequently since a public server is generally visited simultaneously by a high volume of clients all over the world who face very different network conditions. Synchronization of states between the entrance module and the main server module thus become very challenging.}

\hl{To solve this problem,} we introduce a novel stateless salting algorithm. In this algorithm, the entrance module and the main service module do not save any state. Public information is used instead. System timestamp is a desirable choice among the various public information because it changes over time naturally \hl{and is extremely easy to obtain}. We calculate the salt using the timestamp as equation Eq~(\ref{eq11})

\begin{equation}Salt = (System Time - T_0) / X\label{eq11}\end{equation}

This equation is similar to the one-time key generation algorithm specified in RFC 6238\cite{rfc6238}. $T_0$ is the initial value and $X$ is the step size. These two parameters are the same \hl{and kept secret} in the entrance module and the main service module. Even if the attacker speculates about the possible timestamps based on the current time, he cannot obtain the salt because of the confidentiality of $T_0$ and $X$, which further enhances the security of encryption. 

The salting process is described by Eq~(\ref{eq12})

\begin{equation}P\_SA = XOR(Salt, H\_SA)\label{eq12}\end{equation}

In Eq~(\ref{eq12}), P\_SA is described in Eq~(\ref{eq5}).

$XOR()$ is used as the function to add the salt in Eq~(\ref{eq12}). Here $XOR()$ can be replaced by any operations, with the only requirements that\hl{: (1)} the result should be different if the salt changes and \hl{(2)} the operation is reversible. The complexity of $XOR()$ is extremely low, so it is used as the salting function here. This does not introduce any additional security risks. 

The server's verification process is described as the following equations:

\begin{equation}P\_SA = e^{-1}(DA_{65-128}))\label{eq13}\end{equation}

\begin{equation}Salt = XOR(P\_SA, H\_SA)\label{eq14}\end{equation}

\begin{equation}T_s = Salt * X + T_0\label{eq15}\end{equation}

\begin{equation}
Result = (System Time - T_s) \in (0, threshold) ? True : False
\label{eq16}\end{equation}

In the following, we discuss the value of $threshold$ in Eq~(\ref{eq16}). The time required in general occasions and the error redundancy should be considered here. We assume that the timestamp used for encryption in the entrance module is $T_{s1}$ and the system time for verification in the main service module is $T_{s2}$, then the time difference $\Delta T$ can be described by Eq~(\ref{eq17}):

\begin{equation}\Delta T = T_{trans\_en} + T_{trans\_ma} + T_{pro\_cl} + T_{pro\_en} + T_{pro\_ma} + T_{syn} \label{eq17}\end{equation}

In Eq~(\ref{eq17}):

$T_{trans\_en}$ is the transmission delay of the packet from the entrance module to the client; 

$T_{trans\_ma}$ is the transmission delay from the client to the main service module;

$T_{pro\_cl}$ is the processing time of the client between receiving the redirect message and sending the new request to the main service module;

$T_{pro\_en}$ is the processing time of the entrance module from time stamping to sending the message;

$T_{pro\_ma}$ is the processing time of the main service module from receiving the message to verifying the timestamp.

$T_{syn}$ is the system time difference between \hl{the main service module and the entrance module}.

\begin{figure}[htbp]
\centering
\includegraphics[width=12.5cm]{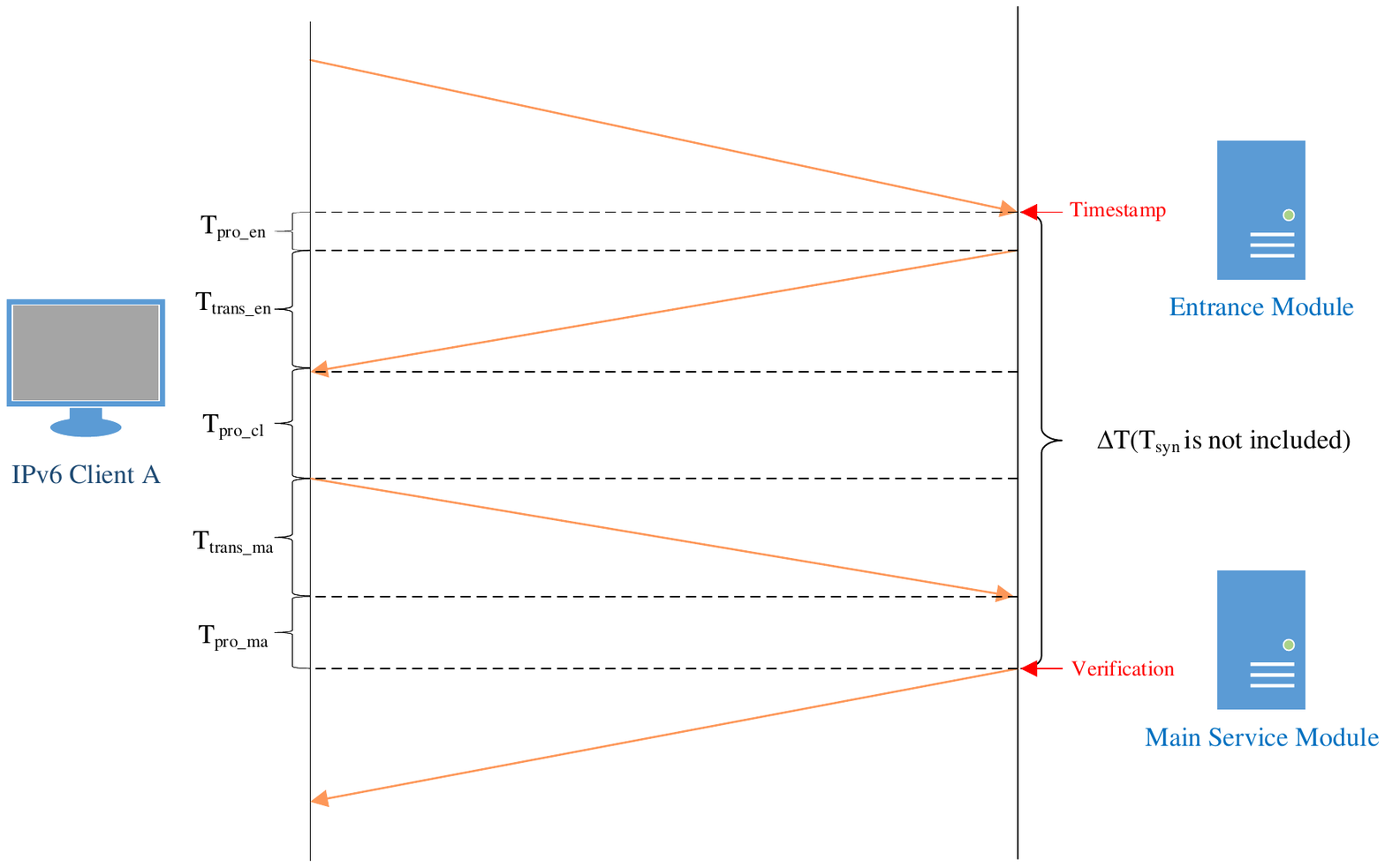}
\caption{{\bf Time \hl{Elapsed} Between Timestamp of Encryption and Verification.} \hl{It consists of two transmission delays (entrance module to client, and client to main service module) and three processing times (entrance module, client, and main service module). Note the system time difference} $T_{syn}$ is not \hl{included}.}
\label{Fig 5}
\end{figure}

It is shown in Fig~\ref{Fig 5}. 

In Eq~(\ref{eq17}), $T_{pro\_en}$ and $T_{pro\_ma}$ are usually negligibly small. \hl{The entrance module and the main service module are both under the control of the server operator, thus the system time difference can be minimized and $T_{syn}$ should also be negligible}. In this case, Eq~(\ref{eq17}) is:

\begin{equation}\Delta T = T_{trans\_en} + T_{trans\_ma} + T_{pro\_cl} \label{eq18}\end{equation}

That is, the threshold should be at least greater than the sum of the delay from \hl{the entrance module to the client}, the delay from the client to the main service module, and the time required by the client to process the redirect message. This usually varies from \hl{several} milliseconds to several seconds. Considering redundancy, a threshold of about 10 seconds is generally appropriate. \hl{A larger threshold brings higher security risks, while a smaller threshold means less redundancy.}

If the synchronization between the entrance module and the main service module is hard, which means $T_{syn}$ in Eq~(\ref{eq17}) cannot be neglected, then the verification function Eq~(\ref{eq16}) should be modified as Eq~(\ref{eq19}):

\begin{equation} Result = (System Time-T_s) \in (-threshold_1, threshold_2) ? True : False \label {eq19}\end{equation}

Here the $-threshold_1$ is negative, because the system time of the main service module \hl{when it performs verification can be earlier than the system time of the entrance module when it performs timestamp. Since the difference in system time synchronization is usually stable, system operators can adjust $threshold_1$ and $threshold_2$ accordingly.} 

However, considering that all modules are controlled by the server owner, time synchronization should not be a challenge. So on most occasions, we can regard $T_{syn}$ negligible. 

\begin{figure}[!ht]
\includegraphics[width=12.5cm]{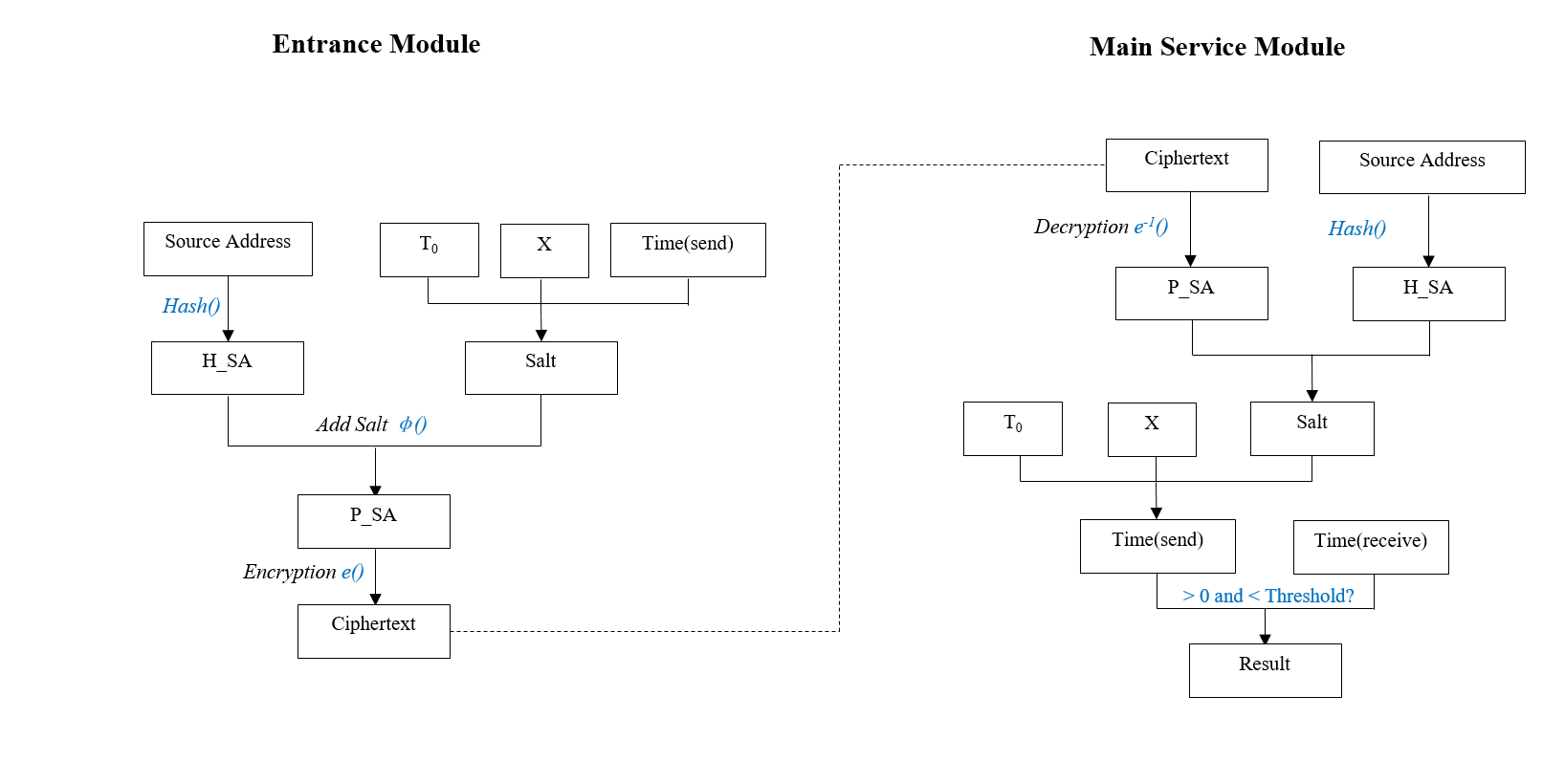}
\caption{{\bf Salting Process.}}
\label{Fig 6}
\end{figure}

The salting algorithm is described in Fig~\ref{Fig 6}.

\subsection*{Prefix Length Consideration}
Generally speaking, according to RFC 4291\cite{rfc4291} and related RFCs, an IPv6 address has a 64-bit prefix and a 64-bit interface identification. Although this is not mandatory, it is consistent with our algorithm. In the above discussion, a 64-bit prefix is allocated to the server which is used for routing, and a 64-bit IID is used to carry the encrypted information. 

\hl{However, our model does not require the prefix length to be 64-bit. }
Assigning a shorter prefix such as /56 is also feasible. 
\hl{The prefix space is used for routing, so} in some cases, the server has to be allocated a longer prefix. For example, the server is built in a /64 subnet, and there are more than one devices in the subnet. In this case, \hl{a /68 or /72 prefix can be assigned.}

\hl{A longer prefix means a shorter suffix, thus less space to carry the encryption information. The suffix length cannot be too short, otherwise, security will be jeopardized. As an extreme example, if the prefix is /120 and the suffix is only 8-bit long, then an attacker can trivially traverse the entire range space of the ciphertext and hit a legal address with only 256 trials. And some encryption algorithms need to be modified to shorten the length of the ciphertext when the suffix is shorter than 64-bit.}

\hl{Further, the prefix length is a flexibly adjustable configuration in our model, because it is only known by the entrance module and the main service module. It is easy to cooperatively adjust the prefix length by the service operator when the network configuration changes.}

\subsection*{\hl{Load Balancing and High Availability}}

In the above discussion, there is only one entrance module and one main service module by default. In fact, our model supports multiple main service modules and/or multiple entrance modules without any modification. 

\begin{figure}[htbp]
\centering
\includegraphics[width=12.5cm]{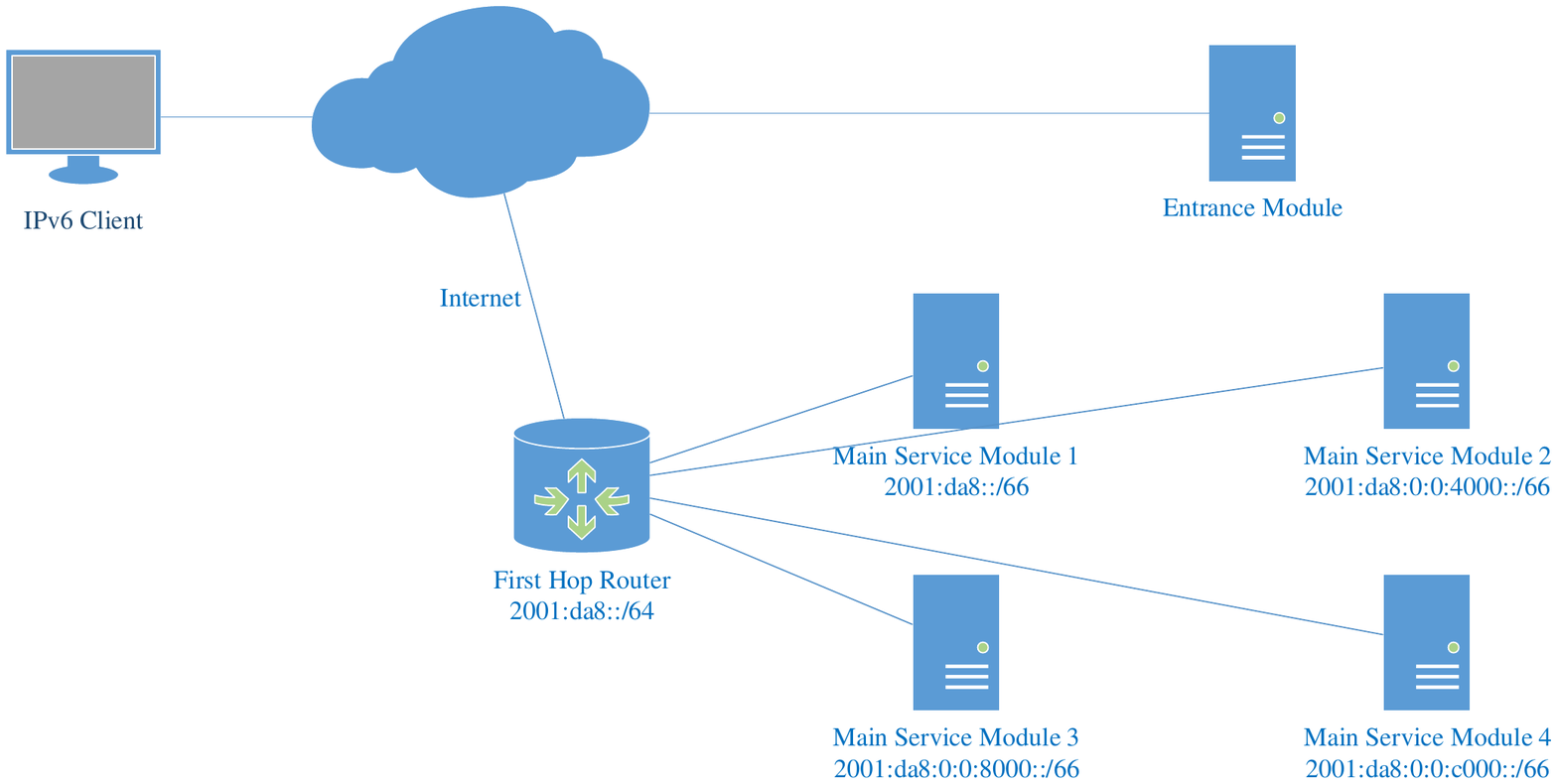}
\caption{{\bf Multiple Main Service Modules and Random Load Balancing.} \hl{In this example, four main service modules jointly serve the /64 prefix of the server. Because the generated address is uniformly distributed and random enough, the load is evenly distributed among them automatically.}}
\label{Fig 7}
\end{figure}

\hl{Multiple main service modules can be deployed smoothly and transparently.} For example, in Fig~\ref{Fig 7}, four main service modules are deployed to jointly serve the /64 address space. In this case, the entire server cluster shares a /64 prefix, but the prefix of each device may have any length longer than /64\hl{, and each can be of different lengths. 
The only requirement is that the prefixes of all the devices should cover the entire /64 address space, otherwise routing will fail and some packets will not be responded. 
Note that overlap is allowed, since the current routing rule of longest prefix matching will ensure proper routing.
The existence of multiple devices, the number of the devices, and the respective prefix length of each device are} all imperceptible to the outside world, and cannot be obtained by any measurement approaches. 
This invisibility obviously helps to protect server security. 

Similarly, multiple entrance modules can also be distributed in different territories and different ISP networks to optimize performance. Because the entrance module \hl{only provides the function of address generation which} is very simple \hl{and stateless, it is even easier} to be arbitrarily stacked in parallel. These entrance modules carry the same key and the same logic, therefore they generate the same address for the same client at the same time. 

\hl{This support for multiple main service modules and/or multiple entrance modules goes beyond capacity expansion. Further, it provides a new and flexible framework to achieve load balancing and high availability related features.

\textbf{Random Load Balancing.} Our model achieve random load balancing automatically without any further modification or configuration. Because the addresses generated by the algorithm in the entrance module are uniformly distributed and random enough, the load is evenly distributed among the suffix space. If the devices all have the same prefix length, then the load is evenly distributed among them. For example in Fig~\ref{Fig 7}, each device has a /66 prefix, so the load will be evenly distributed among these four devices. If the devices have different prefix lengths, then the load is distributed among the devices proportional to the size of the suffix space that is served by each of them. For example in Fig~\ref{Fig 7}, if we delegate 2001:da8::/67, 2001:da8:0:0:2000::/67, 2001:da8:0:0:4000::/66, and 2001:da8:0:0:8000::/65 to the 4 devices respectively, then each will share 1/8, 1/8, 1/4, and 1/2 of the load respectively.

The advantages of the random load balancing feature in our model include:}
\begin{enumerate}
    \item \hl{Simplicity. Random load balancing in our model is completely stateless and requires no scheduling. The devices of the main service module can be arbitrarily stacked, and the load will be automatically distributed among them. To distribute the load in proportion to device capacity is easy. Operators only need to delegate different prefix lengths to the devices and configure routes accordingly.}
    \item \hl{Security. Our address generation algorithm is random. The configuration of devices of the main service module is completely imperceptible and transparent to the outside world. Thus attacks against load balancing become ineffective.}
\end{enumerate}

\begin{figure}[htbp]
\centering
\includegraphics[width=12.5cm]{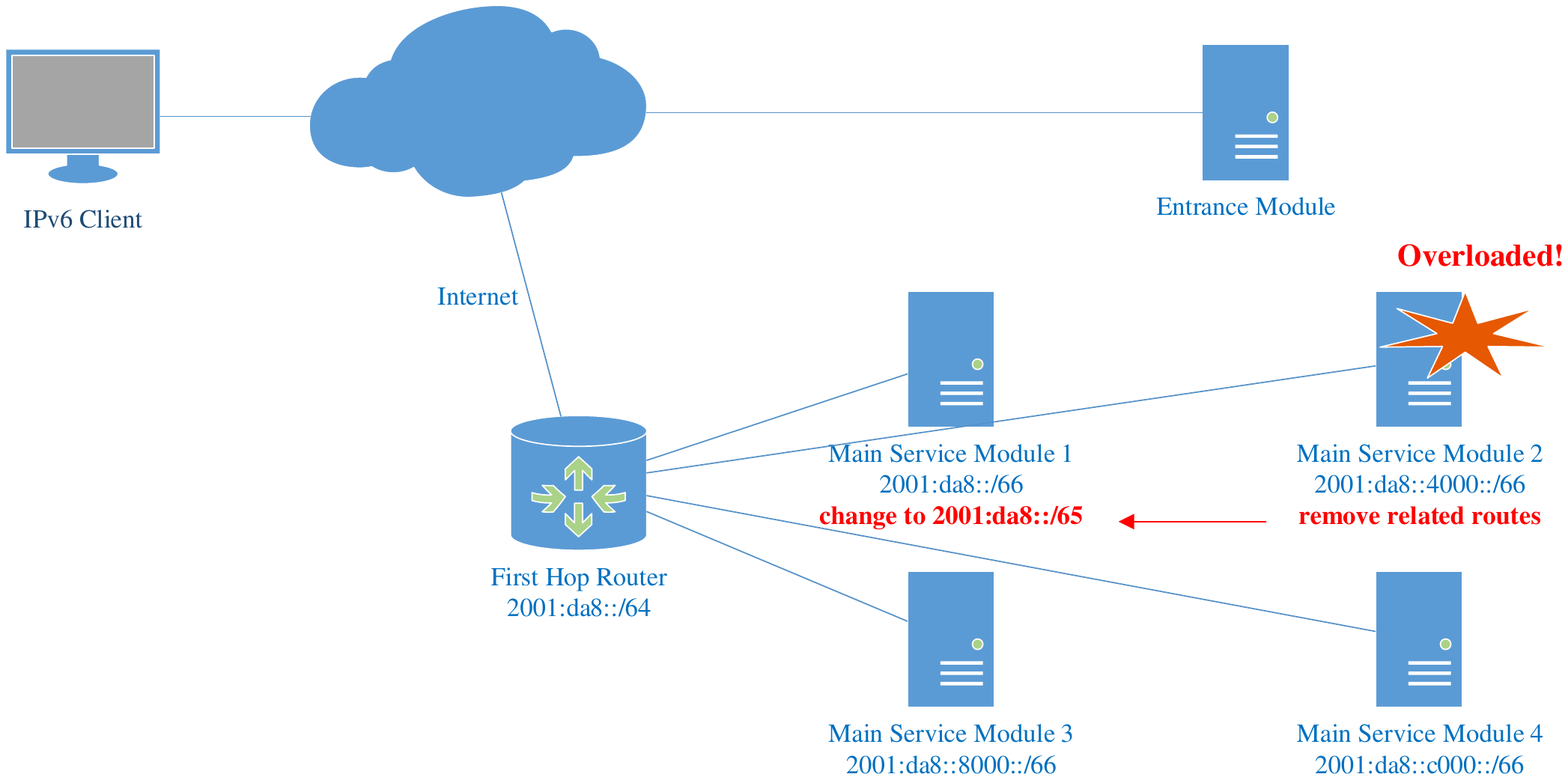}
\caption{{\bf Active-Active High Availability Cluster in Dynamic Load Balancing based on Routing.} \hl{When a device fails, it can go offline by simply redistributing its related prefixes and routes to other devices.}}
\label{Fig 8}
\end{figure}

\hl{However random load balancing has the limitation of lack of control. The load is distributed completely random, thus cannot be controlled according to the status of the devices. Precisely, it is the number of requests instead of the load itself that is distributed, and it is statistically even instead of strictly even at any time. Resources consumed by different connections may vary greatly, and random allocation may not be absolutely uniform. While some devices are temporarily overloaded, others can be idle, resulting in a waste of resources and possible service failure on some devices.

\textbf{Dynamic Load Balancing.} To overcome the above limitation of random load balancing, our model supports dynamic load balancing through two schemes: routing configuration and entrance module strategy.

One way to achieve dynamic load balancing is based on routing, and all the devices of the main service module share one /64 prefix. The delegation of sub-prefixes to each device under this prefix and the related routes can be dynamically configured. In this way, the load of different devices can be adjusted in real-time to balance the utilization of each device. Caution that during the adjustment the sub-prefixes of all the devices should always cover the entire /64 address space.

High availability cluster can be naturally achieved in this scheme, such as active-active cluster and hot standby. When a device is detected to be failing, it can go offline simply by immediately redistributing its related prefixes and routes to other devices, so that the service will not be affected. A self-illustrated example is given in Fig~\ref{Fig 8}.}

\begin{figure}[htbp]
\centering
\includegraphics[width=12.5cm]{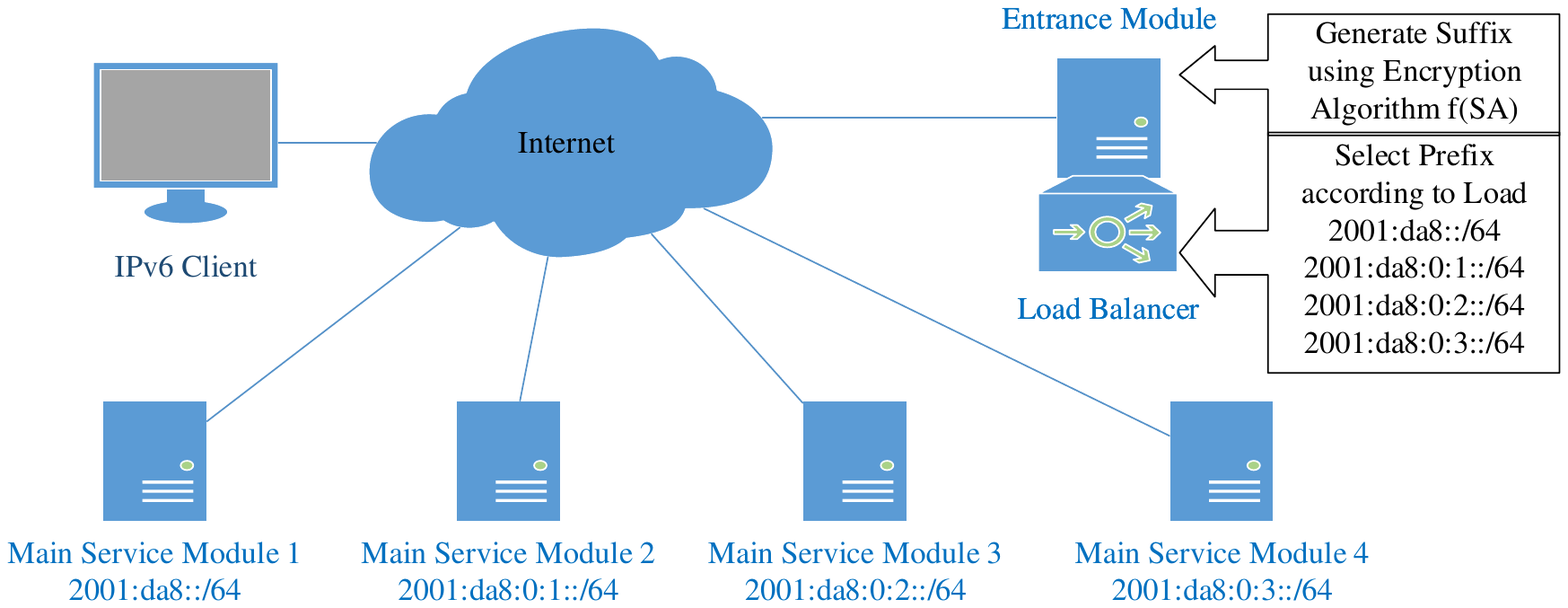}
\caption{{\bf Dynamic Load Balancing based on Entrance Module Strategy.} \hl{ The load balancer resides on entrance module and selects among the prefixes of multiple main service modules according to load.}}
\label{Fig 9}
\end{figure}

\hl{Another way to achieve dynamic load balancing is based on the entrance module strategy, and each of the devices of the main service module is delegated one /64 prefix instead of sharing one /64 prefix. This means the entire server needs a shorter prefix. Nevertheless, for a server that needs dynamic load balancing, a /48 or even shorter prefix is not a problem\cite{rfc6177}. In this case, the entrance module needs to add extra function when generating destination addresses. The 64-bit suffix is still generated as described above, but the 64-bit prefix is no longer the fixed server prefix. Instead, the prefix is selected among the prefixes of the cluster devices using load balancing algorithms, such as round-robin algorithm, least connections algorithm, etc. 

An example is given in Fig~\ref{Fig 9}. The main service module is composed of 4 devices. Each is configured with an /64 prefix, which is 2001:da8::/64, 2001:da8:0:1::/64, 2001:da8:0:2::/64, and 2001:da8:0:3::/64 respectively. Thus the prefix of the entire server is 2001:da8::/62. The load balancer resides in the entrance module, and it selects among these 4 prefixes according to real-time load using a load balancing algorithm. The entrance module generates the suffix using the encryption algorithm $f(SA)$ described in previous sections, then combine it with the selected prefix to generate the destination address.}

\hl{In this case, the devices of the main service module do not need to be connected to the same first-hop router, and can be distributed deployed (Note the difference between Fig~\ref{Fig 9} and Fig~\ref{Fig 8}). Similar to CDN that is based on DNS redirection, the main service modules can be distributed all over the world, and the entrance module is responsible for server selection. The optimal main service module can be selected in a fully controllable manner using information such as geographic location, network capacity, and the ISP of the client.}

\hl{All in all, our model provides a new and flexible framework that naturally supports various load balancing, high availability, and CDN features. For specific network scenarios like data center network, the network topology, structure, and routing configuration can be further optimized. This framework provides great potential for future work.}

\section*{Security Analysis}

The \hl{primary} goal of our model is to prevent the server from being scanned. Therefore, we first discuss \hl{its anti-scanning feature}, then other security features. Finally, we analyze the big data characteristics of \hl{the addresses generated in} our model and \hl{its} ability to resist big data analysis.

\subsection*{Network Scanning}

\hl{As introduced} in the Related Work Section, \hl{recent advances of} IPv6 scanning can be divided into two types:

\begin{enumerate}
\item Scan \hl{by collecting active} IPv6 address records.
\item Scan by generating a hitlist using pattern recognition algorithms.
\end{enumerate}

Scanning by collecting active address records does not pose a threat to our model. 
The main service module is imperceptible to outside. 
If one of the addresses of the main service module \hl{is collected}, as soon as the flow terminates, the address will no \hl{longer} be active.
\hl{This} makes the collected address records \hl{completely useless}.

Scanning by generating hitlists using pattern recognition algorithms does not pose a threat to our model either. 
According to the discussion in Subsection Big Data Analysis, the addresses generated by our algorithm do not have any pattern.
\hl{This} makes scanning through the generated hitlists no different from brute force scanning. 

Another way to scan is to collect active addresses inside the subnet. However, as described in previous sections, the main service module does not use any global unicast address in the subnet. Attackers cannot get any useful addresses from the subnet. Therefore, this approach cannot pose a threat to our model as well. 

As a result, for our model the \hl{above techniques} are no different from brute force scanning, which is generally unworkable in IPv6. 

\begin{figure}[htbp]
\includegraphics[width=12.5cm]{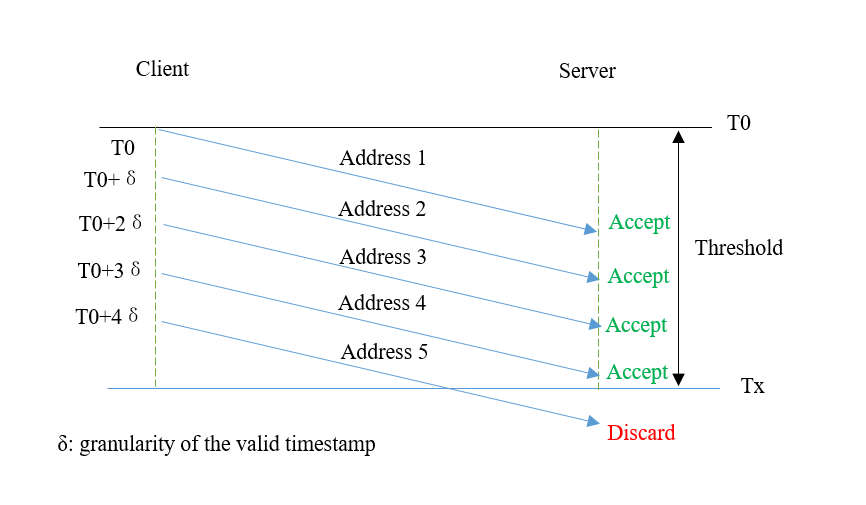}
\caption{{\bf More Than One Legitimate Destination Addresses Can Pass Verification.} \hl{This is due to the $threshold$ introduced in Eq~(\ref{eq16}).}}
\label{Fig 10}
\end{figure}

However, things are complicated by our salting algorithm, which allows more than one destination addresses to pass the verification corresponding to the same source address at the same time. 
For example, if the step size $X$ in Eq~(\ref{eq11}) is 5 milliseconds, and the threshold in Eq~(\ref{eq16}) is 10 seconds, then 2000 legitimate addresses can pass the verification at the same time. 
Assuming that the suffix length of the main service module is $N$, there are $P$ legal timestamps within the threshold, then the probability that a random address can pass the verification is enlarged from $\frac{1}{2^N}$ to $\frac{P}{2^N}$. 
Fig~\ref{Fig 10} shows an illustration, where all addresses from Address 1 to Address 4 can pass the verification. 

\hl{With} the \hl{same} scanning efficiency of Zmap under IPv4, which takes 45 minutes to scan the $2^{32}$ address space, the expected time $T$ of the scanning will be

\begin{equation}T=\frac{3*2^{N-32}}{4P}\label{eq20-1}\end{equation}

To protect the server from being scanned, $T$ should be long enough, for example, longer than 1 year. Then P and N should satisfy Eq~(\ref{eq20})

\begin{equation}N - log_2{P} \ge 46\label{eq20}\end{equation}

Eq~(\ref{eq20}) is easy to satisfy. For example, if a threshold of 10 seconds and a step size of 1ms are used, the expected scanning time $T$ is about 9 years, which is too long to be worried about. Therefore, \hl{the complexity introduced by the salting algorithm does not harm the anti-scanning feature of our model}.

\hl{Our above discussion focus on preventing the main service module from being scanned. 
Note that the entrance module is inevitably exposed to scanning, because some sort of entry address must be configured into the DNS system. 
However, first, scanning the entrance module does not pose a threat to the main server, because the address that provides the main service has been separated from the entrance module. 
Second, unnecessary exposure is eliminated through this separation, and the main service module, as the main body of the server, is well protected.
Third, the entrance module is somehow similar to a bastion host. It provides no business logic, and the consequences of it being scanned are much more limited. Thus risks are isolated and controlled to a smaller scope.}

\subsection*{DoS Attack}
DoS attack is one of the major threats faced by high-profile public servers. We discuss two \hl{typical} types of DoS attack here: TCP Syn-Flood and UDP Flood.

Syn-Flood does not pose a threat to our model. 
All messages received by the main server are verified, and the ones with inappropriate source address-destination address pairs will be discarded. 
This means that the server does not need to send SYN+ACK messages, nor allocate CPU time or memory to maintain a queue waiting for subsequent packets. 

Similarly, for UDP Floods, all the illegal UDP packets will be dropped immediately, and no resources will be allocated. 
Therefore, neither Syn-Flood nor UDP Flood poses a threat to our model.

Admittedly, the mitigation of DoS attacks is also limited. Our model cannot prevent the types of attacks that do not target the server directly, such as bandwidth attacks. \hl{Nor can it prevent attacks in which each of the malicious requests first gets a legitimate address from the entrance module. In this case, an intrusion detection system can be deployed on the entrance module to detect and respond to abnormal traffic.}

Again, the entrance module is still exposed to DoS attacks. 
However, the entrance provides limited service with little traffic load, so its ability to withstand DoS attacks is much higher. 
And its function is very simple, so it is easy to be duplicated, stacked, and distributed deployed, which further enhances its resistance to DoS attacks.

\subsection*{Application Vulnerability Attack}

To prevent application vulnerability attacks generally requires software and hardware updates in time. 
However, our model can mitigate this threat by applying stricter authentication in the entrance module. 
The entrance module can be configured with authentication or admission strategies, so that only an authorized user can get the legitimate address, and the other requests are discarded. 
In this way, adversaries cannot even obtain the address of the main service module, let alone exploiting application or operating system vulnerabilities to launch attacks such as SQL injection attacks.

\subsection*{Replay Attack and Session Hijack}
A salting algorithm is introduced in our model to prevent replay attacks. The destination address generated by the entrance module is different each time. If an attacker launches replay attacks using previously intercepted packets, the address will already become illegal.

\hl{However, there is a complication introduced by the $threshold$ in Eq~(\ref{eq16}). 
Though the entrance module dutifully generates a different address each time, a given address can pass the verification of the main service module in a time window of length $threshold$.
Thus if an attacker intercepts a packet and launches replay attacks fast enough within $threshold$, the address will remain valid and pass the verification. 

It can be solved by requiring the main service module to cache the addresses used in connections that have just ended within a time window of length $threshold$, and to reject connection requests with destination addresses that have been used recently. 
However, this caching is memory intensive for a high-traffic server. 
New vulnerabilities like cache exhaustion attack can be introduced, though the attacker needs to contact the entrance module first to get a huge amount of legitimate addresses.
This strategy, accompanied by an intrusion detection system deployed on the entry module, can be an option for servers with high security requirements but not too much traffic.}

Similar to fast replay attacks, the attacker can monitor the connection and launch session hijacking attacks. This is because verification is conducted only once for each flow in our model for performance considerations. Once the connection is established, the client and the main service module communicate directly without verification, which can be taken advantage of.

\hl{These two kinds of attacks can both} be prevented using encryption at higher layers. 
Considering that our model is a network layer model, it is orthogonal thus compatible with encryption protocols at the transport or application layer, such as TLS\cite{tls}, DTLS\cite{dtls}, and HTTPs\cite{https}.
These protocols can encrypt the packet payload and invalidate replay attacks and session hijacking attacks.

\subsection*{Key Cracking}
For breaking symmetric encryption, the plaintext-ciphertext pair needs to be obtained to guess the key.
Even if an attacker intercepts the messages, the plaintext of the function $e()$ in Eq~(\ref{eq6}) is confidential and cannot be inferred from the source address.
This is because the salting algorithm in our model makes the plaintext time-varying.
The precise timestamp is difficult to get, let alone the parameters $T_0$ and $X$ are confidential.
As a result, although DES is considered insecure in the modern Internet, it has a sufficient security level in our model. 


\subsection*{ND related Attack}

ND attacks are unique in IPv6 because the ND protocol\cite{ND} is designed to replace the ARP\cite{ARP} protocol of IPv4. 
There are many kinds of ND related attacks, which can be roughly divided into DAD attacks\cite{b1,b2}, spoofing attacks (such as RA spoofing, malicious redirection), and cache exhaustion attacks.

Our model allocates a prefix to the main service module instead of an address, and there is only one device under this prefix. A lot of ND packets used for neighbor communication are no longer needed. Thus related threats are eliminated, including rouge NA and NS messages and DAD DoS Attacks\cite{DADattack}. 

On the other hand, ND messages are still used in the prefix allocating process by some mechanisms such as DHCP-PD. Thus RA spoofing attacks still pose a threat.
And link-local addresses are used to communicate with the first-hop router in the subnet in our model, so ND attacks on the link-local addresses are still effective.
Meanwhile, the entrance module still has a fixed address that can be attacked. 

Therefore, although our model protects the main service module from a lot of ND attacks, it is recommended to deploy ND-related security policies such as SEND\cite{send1,send2}, RA-guard\cite{raguard1,raguard2}.

\subsection*{Big Data Analysis}
In our model, one concern is that the attacker obtains a lot of source addresses-destination addresses pairs by intercepting a large number of network traffic data, and learns patterns of the generated addresses to launch attacks and scans. However, if the generated addresses have sufficiently good statistical characteristics, that is, random and evenly distributed enough, attackers cannot crack, scan, or attack by big data analysis. 

\begin{figure}[!h]
\centering
\includegraphics[width=12.5cm]{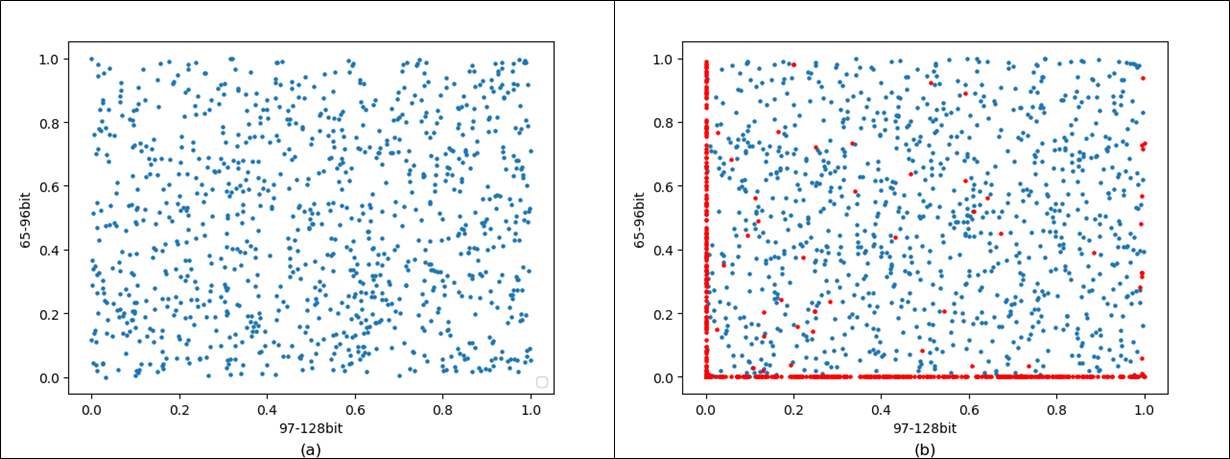}
\caption{{ \bf Distribution of the Generated Suffixes.} 
(a) is the result of Group 1\hl{, where 1000 destination addresses are generated using one fixed source address at different times}. (b) is the result of Group 2, \hl{where 1000 destination addresses (blue points) are generated using 1000 source addresses (red points) collected in real-world campus network traffic. The generated suffixes are evenly distributed in both groups.}}
\label{Fig 11}
\end{figure}

\hl{We use simulations to demonstrate that the destination addresses generated using our algorithm have good statistical characteristics. In the simulation, we generate a large number of destination address suffixes then analyze them. Our simulation is conducted in 2 groups, and the results are shown in Fig~\ref{Fig 11}.}

\hl{The figure is plotted in the following manner. 
Since our algorithm generates address suffixes, we show the scatter plot of the generated suffixes.
The longitudinal coordinate of each point is determined by the 65-96 bits of the address.
For example, if the 65-96 bits are 0000:0000, then the vertical coordinate is 0; and if the 65-96 bits are FFFF:FFFF, then the vertical coordinate is 1.
Similarly, the horizontal coordinate of a point is determined by the 97-128 bits of the address in the same way. }

\hl{In Group 1, we use a fixed address as the source address, and generate 1000 destination address suffixes using it. The salt used in the process naturally varies over time, so the generated addresses are different. The distribution of the generated suffixes is shown in Fig~\ref{Fig 11}(a). It shows that suffixes generated using the same source address over time are sufficiently evenly distributed and have no obvious pattern.}

\hl{In Group 2, we collect 1000 active addresses in real-world network traffic from Tsinghua Campus Network.
Then we use these 1000 source addresses to generate 1000 destination address suffixes. 
The results are shown in Fig~\ref{Fig 11}(b). 
The red points are the suffixes of the source addresses, while the blue points are the suffixes of the generated addresses. 
An interesting observation is that the collected IPv6 addresses have an obvious pattern. 
A huge number of addresses set all of the 65-96 bits to zero, while another huge number of addresses set all of the 97-128 bits to zero.
Nevertheless, the suffixes generated by our algorithm using these collected addresses are still evenly distributed.}

Furthermore, to evaluate the randomness of the generated addresses, we calculated the entropy of them. This approach is often used to evaluate the randomness of IP addresses\cite{b27,b28,b29,b30}. 
The entropy is calculated for each nybble (a nybble is 4-bit), so there are 16 entropy values in total \hl{for} a 64-bit suffix. 
\hl{In probability theory, for a discrete random variable $X$ with possible values $\{x_1, ..., x_k\}$ and probability mass function $P(X)$, entropy is defined as}
\begin{equation}
    H(X) = -\sum_{i = 1}^{k}P(x_i)logP(x_i)
\end{equation}

\hl{For each nybble in our case, there are 16 possible hex characters. From probability theory we know that $H(X)$ is maximum if $P(X)$ is uniform, so here the maximum entropy is log16. For the k-th nybble, if character $c_i$ occurs $N_i$ times (assume there are N samples totally), the entropy normalized using the maximum entropy is} 

\begin{equation} 
e_k = -\frac{1}{4}\sum_{i=1}^{16} \frac{N_i}{N}log(\frac{N_i}{N}) 
\label{eq25}\end{equation}

\begin{figure}[htbp]
\centering
\includegraphics[width=12.5cm]{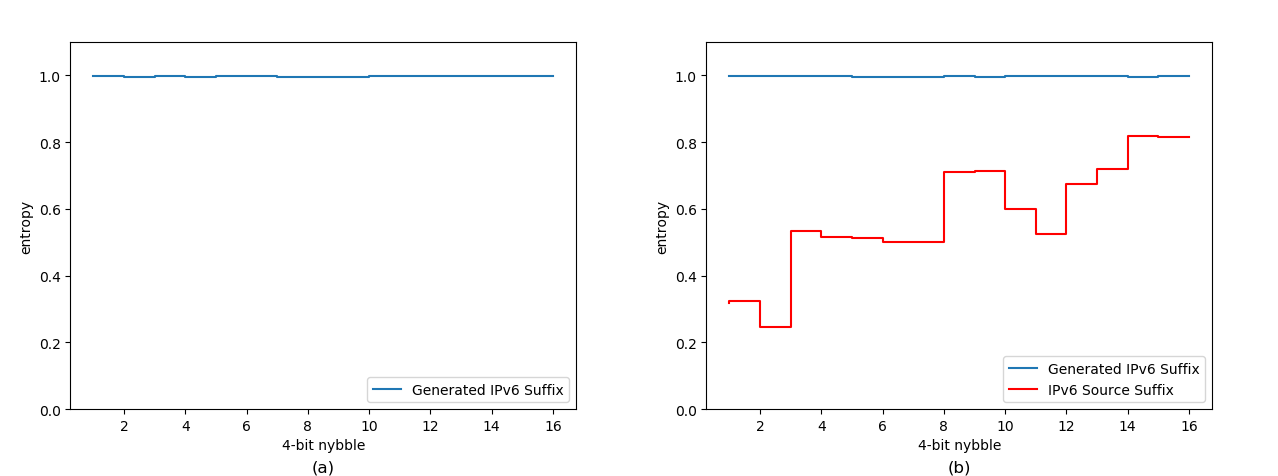}
\caption{{\bf Entropy of the Generated Suffixes.}
\hl{(a) is the result of Group 1, where 1000 destination addresses are generated using one fixed source address at different times. (b) is the result of Group 2, where 1000 destination addresses are generated using 1000 source addresses collected in real-world campus network traffic. Blue line shows the entropy of the suffixes of the generated destination addresses, while red line shows the entropy of the suffixes of the source addresses. In both groups, the entropy of the generated suffixes is almost 1, which is the max entropy. The generated address suffixes are sufficiently random in all bits.}}
\label{Fig 12}
\end{figure}

The results are shown in Fig~\ref{Fig 12}. 
\hl{Fig~\ref{Fig 12}(a) shows the results of Group 1, and Fig~\ref{Fig 12}(b) shows the results of Group 2. 
The red line is the entropy of the source address suffixes, while the blue line is the entropy of the generated destination address suffixes. 
It is interesting yet reasonable that the collected addresses are not so random, especially for higher nybbles.
In both groups, the entropy of the generated suffixes for each nybble is almost 1. 
Since the entropy is normalized using the maximum entropy, this means that no matter the source addresses are random or not, the generated address suffixes are sufficiently random in all bits.}

\hl{In short, our model can withstand big data analysis attacks launched by obtaining and analyzing large volumes of traffic data.}

\section*{Prototype Implementation and Experiments}
In this section, we introduce our prototype implementation and experiments based on the prototype.

\subsection*{Prototype Implementation}
Our prototype is implemented based on Linux, specifically, Raspbian OS based on Debian. We make modifications to the NetFilter Linux kernel module to implement the mechanism. In our prototype, the following features are implemented:
\begin{enumerate}
\item The server is divided into two modules. The entrance module is configured with an IPv6 address while the main service module is assigned a prefix using DHCP-PD. The main service module listens on all the addresses under the prefix.
\item The entrance module and the main service module save the same key. When the entrance module receives a request, it generates an address under the main service module prefix and redirects the request to that address. When the main service module receives a request, it conducts verification on the flow.
\item The encryption and the verification use DES as e() in Eq~(\ref{eq6}), and the salt is added as Eq~(\ref{eq11}) to prevent replay attacks.
\end{enumerate}

The algorithm implemented in the prototype is described as follows. The encryption algorithm executed in the entrance module is Algorithm 1, while the verification algorithm executed in the main service module is Algorithm 2.

\begin{algorithm}
\caption{Encryption Algorithm Executed by the Entrance module}
\label{alg1}
{\bf Input:}
SourceAddress,Key,Prefix,T0,X \\
{\bf Output:}
DestinationAddress

\begin{algorithmic}[1]
\STATE H\_SA = DJB(SourceAddress)
\STATE T\_current = time.currenttime()
\STATE Salt = (T\_current - T0) / X
\STATE P\_SA = XOR(Salt, H\_SA)
\STATE Suffix = DES(P\_SA,Key)
\STATE DestinationAddress = strcat(Prefix, Suffix)
\RETURN{DestinationAddress}
\end{algorithmic}
\end{algorithm}

\begin{algorithm}
\caption{Verification Algorithm Executed by the Main Service module}
\label{alg2}
{\bf Input:}
SourceAddress,DestinationAddress,Key,T0,X,T\_threshold \\
{\bf Output:}
True/False
\begin{algorithmic}[1]
\STATE H\_SA = DJB(SourceAddress)
\STATE Suffix = DestinationAddress[64:128]
\STATE P\_SA = DES$^{-1}$(Suffix, Key)
\STATE T\_current = time.currenttime()
\STATE Salt = XOR(H\_SA, P\_SA)
\STATE T\_send = Salt*X + T0
\STATE T\_delta = T\_current - T\_send
\IF{T\_delta < T\_threshold \textbf{and} T\_delta > 0}
\RETURN{True}
\ENDIF
\RETURN{False}
\end{algorithmic}
\end{algorithm}

\subsection*{Experiment Environment}
Our experiment is built on a /48 subnet in CERNET (China Education and Research Network). The entrance module and the main service module are configured in the same subnet. We use DHCP-PD to allocate a prefix to the main service module while the addresses of the client and the entrance module are configured manually. We use Kea for the DHCP server.

We use Raspberry Pis for all the devices in the experiment, and the operating system of each Pi is Raspbian, which comes with the Raspberry Pi and is based on Debian. 

\begin{figure}[!h]
\centering
\includegraphics[width=12.5cm]{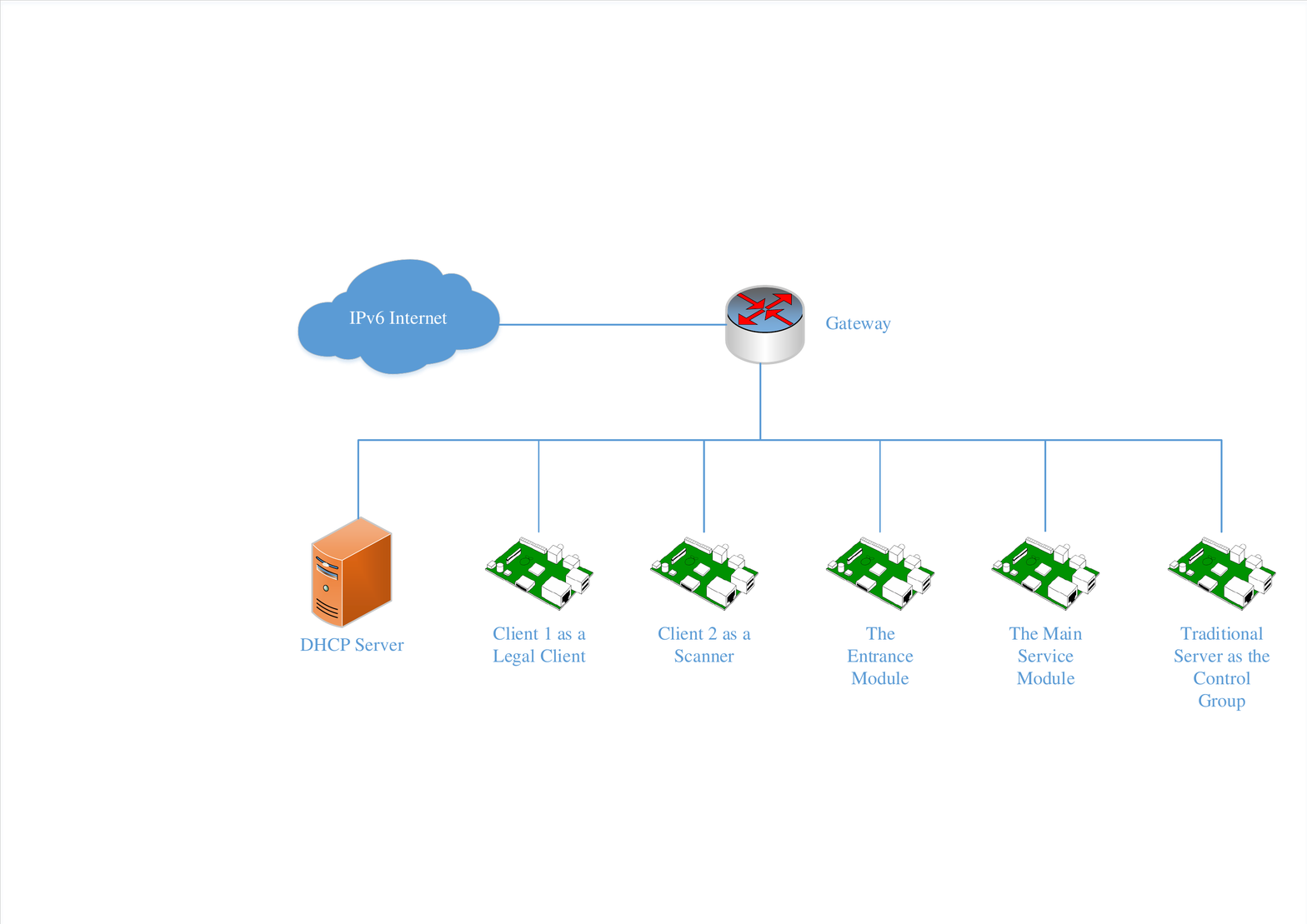}
\caption{{\bf Topology of the Experiment Subnet.}}
\label{Fig 13}
\end{figure}

The experiment topology is described in Fig~\ref{Fig 13}.

Five devices are configured in the experiment subnet. One is a legal client, one is a scanner, one is the entrance module of the server, and one is the main service module. To offer a baseline for the experiment, we configure the last one a traditional server as the control group. 

\subsection*{Experiment on Defense of Scanning}
In the experiment, the server is configured as \hl{a public HTTP web server based on the apache and Django framework. The demo web page displays the source address and destination address of the visit for the demonstration.} We use \hl{a} client to access the server through the entrance module first, and the result shows that the client can visit the main server smoothly, while the client cannot get any response if it tries to visit the main server directly. This shows that the main server cannot be perceived by the outside world directly.

To test the scanning defense feature of the main server, a client is used as the attacker. Three types of scans are tested in the experiment. First, we perform a 100-hour brute-force scan on the main service module. The result shows that the scan does not hit any address.

Then, we generate 1 million addresses under the /64 prefix using Entropy/IP and 6gen, and use them as the hitlists to scan the server. The scan does not hit any address as well.

Finally, ND information for other devices in the subnet is collected to compose a hitlist to conduct the \hl{scan. Similarly}, no address can be accessed as well. Therefore, our model can prevent the main service module from being scanned by existing IPv6 scanning approaches.

\subsection*{Experiment on Performance}
\hl{We conduct two sets of experiments on performance in this subsection. First on the delay in the connection establishment phase, then on the RTT, bandwidth, and jitter after the connection is established.}

Because our model introduces additional overhead \hl{only} in the connection establishment phase which influences the delay most, we first conduct experiments on the delay in the connection establishment phase. \hl{We use Wireshark to monitor the delay.} Compared with the control group, the extra delay \hl{introduced by our model equals the time elapses between the entrance module receiving the first packet and the main server module receiving the first packet. This value} can be described by Eq~(\ref{eq22}). 

\begin{equation}T_{add} = T_{trans\_en} + T_{trans\_ma} + T_{process\_cl} + T_{encryption\_en} + T_{verification\_ma}\label{eq22}\end{equation}
 
In Eq~(\ref{eq22}), $T_{trans\_en}$ is the transmission delay \hl{from} the entrance module \hl{to} the client; $T_{trans\_ma}$ is the transmission delay \hl{from} the client \hl{to} the main service module; $T_{process\_cl}$ is the processing time from the client receiving the redirect message to the client sending the new request to the main server; $T_{encryption\_en}$ is the \hl{processing time from the entrance module receiving the request to it sending the redirect message; while $T_{verification\_ma}$ is the processing time from the main service module receiving the request to it completing the verification. We test these five items separately to analyze to which degree each of them influences the delay.}
 
\hl{First, $T_{trans\_en}$ and $T_{trans\_ma}$ typically vary from several to hundreds of milliseconds. The one-way transmission delay measured in our experiment environment is very small (several milliseconds), but this value is greatly affected by the specific network environment of each client, so our results are not representative and we will not display it here.} However, considering that the additional cost caused by two one-way delays is usually a small number (generally at most a few hundred milliseconds) comparing with the total access time cost, \hl{it may not cause a user-perceivable impact.}

\begin{figure}[!h]
\centering
\includegraphics[width=9cm]{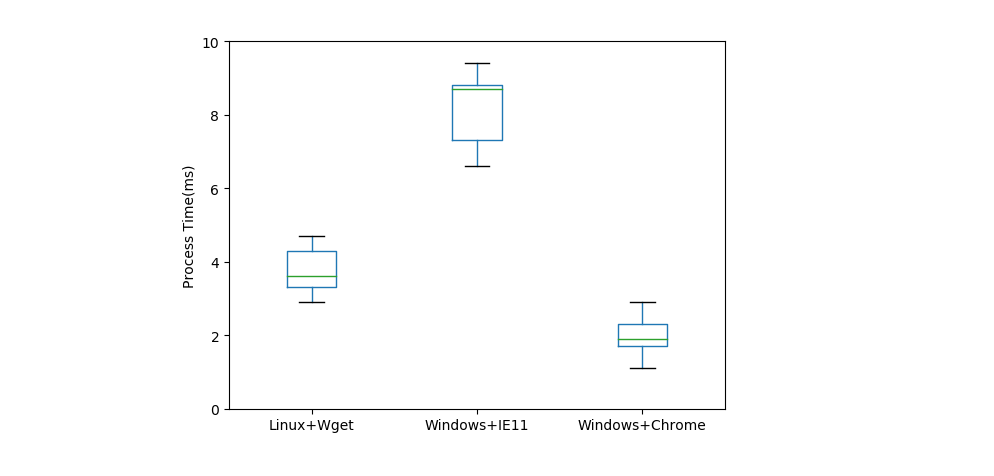}
\caption{{\bf $T_{process\_cl}$ with Different \hl{OSes} and Browsers.} 
\hl{We measure the process time of the client from receiving the redirect message to it sending the new request to the main server. It is greatly affected by the environment, but will not exceeds 10 milliseconds.}}
\label{Fig 14}
\end{figure}

 
\hl{Second, we test} the effect of $T_{process\_cl}$. $T_{process\_cl}$ is dependent on the hardware and the application of the client. We conducted the experiments in three groups \hl{of client settings}: Group (a) uses Linux and Wget, Group (b) uses Windows and IE 11 Browser, and Group (c) uses Windows and Chrome Browser. 
\hl{Each client launches access to our prototype website and $T_{process\_cl}$ is measured in each test using tcpdump/Wireshark.} Each group launches access to our prototype website for five times, with an hour interval between each. The result is shown in Fig~\ref{Fig 14}. 


From Fig~\ref{Fig 14}, we can see that $T_{process\_cl}$ is \hl{greatly affected by different hardware and software environment}, but it is less than 10 milliseconds in all groups. It means that $T_{process\_cl}$ brings little additional \hl{delay and does not influence user experience at all.}

\begin{figure}[!h]
\centering
\includegraphics[width=9cm]{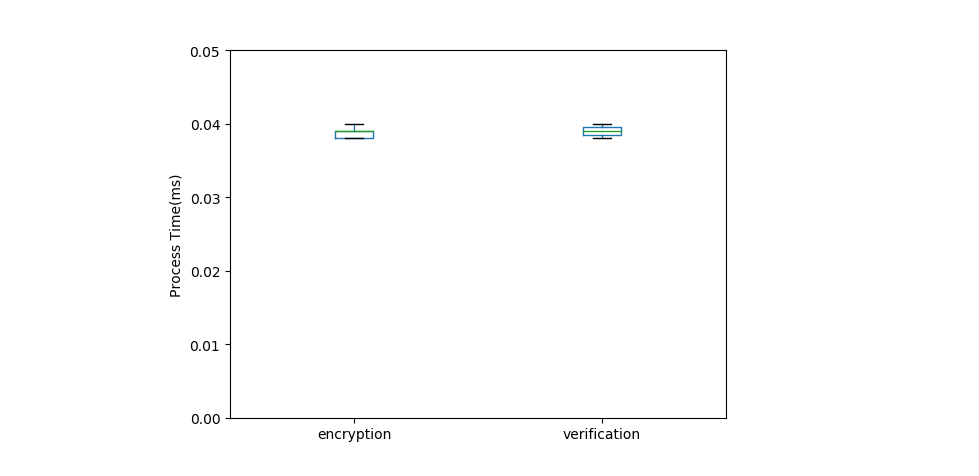}
\caption{{\bf Encryption Time and Verification Time.} \hl{The time cost is less than 0.05 milliseconds, and its influence to server performance is negligible.}}
\label{Fig 15}
\end{figure}

\hl{Finally, $T_{encryption\_en}$ and $T_{verification\_ma}$ represent the execution time of the encryption and verification process, respectively.} 
We conduct experiments on the encryption and verification time cost. The experiment is repeated 10 times. 
The result is shown in Fig~\ref{Fig 15}.

From Fig~\ref{Fig 15}, we can see that \hl{the execution time of a single encryption or verification operation does not exceed 0.05 milliseconds}, which is negligible to the delay and will not affect the server \hl{performance}. 

\hl{In summary, the additional delay brought by our model is determined by the two one-way transmission delay, while the sum of the other time overhead will not exceed several milliseconds. The total additional delay is acceptable and will not affect user experience.}

\begin{figure}[htbp]
\centering
\includegraphics[width=12.5cm]{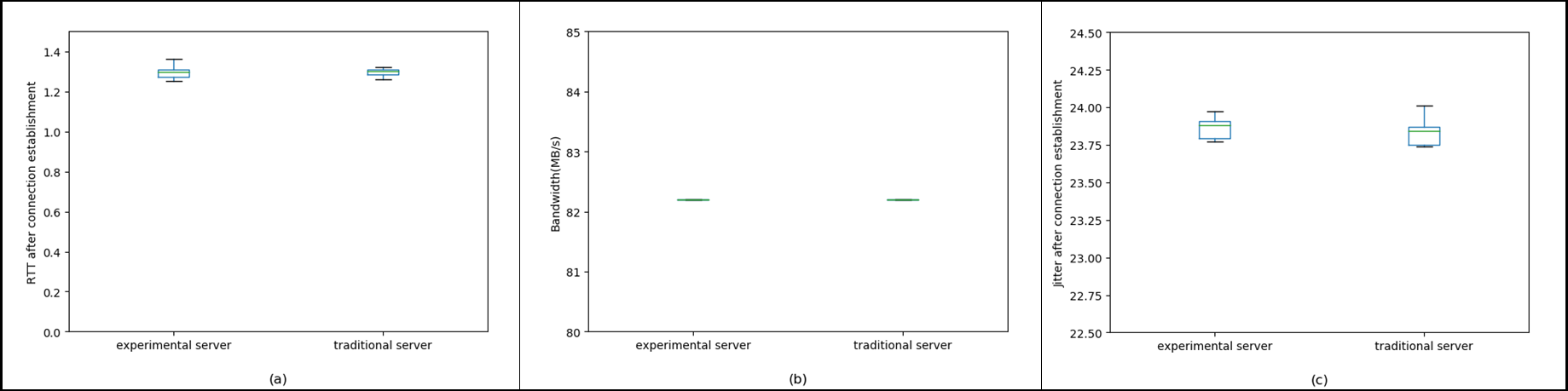}
\caption{{\bf RTT, Bandwidth, and Jitter Between Client and Server after Connection Establishment.} \hl{There is no performance difference between our model and a traditional server. Our model brings no additional cost once the connection is established.}}
\label{Fig 16}
\end{figure}

In order to test the impact of our model on network performance after the connection is established, we conduct experiments on the RTT, bandwidth, and jitter between the client and the server. The result is shown in Fig~\ref{Fig 16}. \hl{Our model does not bring any additional cost in RTT, bandwidth, or jitter once the connection is established.}

\section*{Discussion}

\subsection*{Compatibility, Deployability, and Evolvability}
Our model has good compatibility with the current Internet.
A client \hl{agnostic of our model} can visit the server smoothly. 
\hl{And none of} the other participants of the Internet need any modification, such as ISP or DNS servers.
Moreover, the modification is limited within the network layer, thus the model transparently supports protocols of the transportation layer and application layer. 
All in all, our model requires no transition mechanism, is simple, and can be easily deployed.

\hl{Not only can our model} provide a new perspective on many security and functional issues \hl{including the} scanning problem, but also it provides a new and flexible framework for public servers. Various strategies can be formulated based on \hl{this framework, such as user authentication on the entrance module}, distributed deployment, and \hl{load balancing and high availability of the server cluster}, etc. We believe that various exciting models can be proposed based on this framework. Therefore, our model has good scalability \hl{and potential}. 


\subsection*{Consideration on IPv6 Address Space}

\hl{Is it too extravagant that we assign an entire prefix to one server in our model?}
Although the number of prefixes that can be allocated is much fewer than the number of addresses in the IPv6 address space, the worry that the IPv6 addresses will be exhausted is unnecessary.
All addresses under 2000::/3 are global unicast addresses. 
There are 8.6 billion global unicast /36 prefixes in total, which means that everyone in the world can be allocated a /36 prefix, not to mention that there are 268 million /64 prefixes under a single /36 prefix. 
Even if a server is allocated a /56 prefix or even a /48 prefix, the current IPv6 address space is still enough. 
Therefore, assigning a prefix to each server will not exhaust the IPv6 address space.


\subsection*{Consideration on IPv4}

Is our model available for IPv4? Theoretically, yes. 
In IPv4, the server can also be divided into an entrance module and a main service module, the main service module can also be configured with a block of addresses while the entrance module can also generate a verifiable address using \hl{encryption} algorithm and provide redirection. 
However, there are two problems in IPv4:
\begin{enumerate}
\item In IPv6, we use a 64-bit suffix length to carry the ciphertext. However, in IPv4, the entire address space is only 32-bit. It cannot provide sufficient security level in encryption, resulting in the encryption being easily cracked.
\item Different from the \hl{high} redundancy of IPv6 address space, IPv4 addresses are very scarce. In IPv4, to allocate a block of addresses to one server is \hl{wasteful and literally too expensive\footnote{https://ipv4marketgroup.com/ipv4-pricing/}}. Therefore, it is \hl{unrealistic}.
\end{enumerate}

Therefore, our model is only recommended to \hl{be used} in IPv6.


\section*{Conclusion}
In this paper, a novel Internet server model named addressless server is introduced. 
This model separates the entrance module from the main service module, and \hl{uses prefix delegation mechanism to} allocate an IPv6 prefix instead of an IPv6 address to the main service module. 
When a user sends a request to the server, the entrance module generates an address \hl{under the main service module prefix by encrypting} the user address, then redirects the request to the generated address. 
A time-varying salt is added in the encryption \hl{process} to make the address different in each connection. 
The main service module verifies each request received and drops all packets that failed verification. 
In this way, our model can prevent the main server from being scanned and other security threats such as DoS attacks and replay attacks. 
 
Our model takes advantage of the massive IPv6 address space to hide the address in use, and incorporates encryption into IPv6 addresses, which can fully utilize the redundant space of IPv6 address. 
By allocating a prefix to the main service module, our model eliminates the one-to-one correspondence between the server and the IP address.
\hl{Our model not only shields the server from scanning, but also establishes a new network framework for servers that supports desirable features such as load balancing, active-active cluster, and CDN conveniently.}

 
We implement a prototype of the addressless server and conduct simulations and experiments based on it.
The results show that users can access the server smoothly while scan traffic cannot get any response.
The addresses generated by the algorithm are sufficiently random and uniformly distributed, which prevents attackers from using big data analysis to scan the servers or crack the keys.
\hl{Our experiments on the performance show that only during the establishment of the connection does it bring additional delay}, and it does not affect user experience. Once the connection is established, our model will not affect delay, bandwidth, or jitter.

%
%
%




\end{document}